\documentclass[prx,superscriptaddress,twocolumn,footinbib]{revtex4-1}
\pdfoutput=1
\usepackage[colorlinks=true,urlcolor=blue]{hyperref}
\usepackage[normalem]{ulem}
\usepackage{mathtools}
\usepackage{amsfonts}
\usepackage{graphicx}
\usepackage{stmaryrd}
\usepackage{amsmath}
\usepackage{amssymb}
\usepackage{lineno}
\usepackage{xcolor}
\usepackage{color}
\usepackage{bbold}
\usepackage{bm}

\newcommand{\stkout}[1]{\ifmmode\text{\sout{\ensuremath{#1}}}\else\sout{#1}\fi}

\renewcommand{\Re}{\mathfrak{R}}
\renewcommand{\Im}{\mathfrak{I}}

\DeclareMathOperator{\Ei}{Ei}
\DeclareMathOperator{\tr}{tr}
\DeclareMathOperator{\Arg}{Arg}
\DeclareMathOperator{\sign}{sign}
\DeclareMathOperator{\arcsinh}{arcsinh}

\newcommand{\traceless}[1]{\left\llbracket #1 \right\rrbracket}
\newcommand{\OPeq}{|\Psi_{0}|}
\newcommand{\OP}{|\Psi|}
\newcommand{\deps}{\dot{\epsilon}}
\newcommand{\Hp}{\mathfrak{H}_{p}}
\newcommand{\hp}{\chi}
\newcommand{\re}{{\rm Re}}
\newcommand{\er}{{\rm Er}}
\newcommand{\D}{\mathcal{D}}
\newcommand{\De}{\D_{\rm eff}}
\newcommand{\Ge}{\gamma_{\rm eff}}

\newcommand{\U}{\mathfrak{U}}
\newcommand{\ls}{\ell_{\rm s}}
\newcommand{\Rs}{\mathcal{R}_{\rm s}}

\definecolor{red}{rgb}{0.75,0,0}
\definecolor{blue}{rgb}{0,0,0.75}
\definecolor{green}{rgb}{0,0.5,0}

\newcommand{\red}[1]{{\color{red} #1}}

\newcommand{\green}[1]{{\color{green} #1}}

\newcommand{\john}[1]{\red{[{\bf JT:} #1]}}

\newcommand{\jtr}[1]{\john{I removed an equation here}}

\begin{document}
	
%\linenumbers

\title{Hydrodynamic theory of $p-$atic liquid crystals}

\author{Luca Giomi}
\email{giomi@lorentz.leidenuniv.nl}
\affiliation{Instituut-Lorentz, Universiteit Leiden, P.O. Box 9506, 2300 RA Leiden, The Netherlands}

\author{John Toner}
\affiliation{Department of Physics and Institute of Theoretical Science,  University of Oregon, Eugene, Oregon 97403, USA}

\author{Niladri Sarkar}
\affiliation{Instituut-Lorentz, Universiteit Leiden, P.O. Box 9506, 2300 RA Leiden, The Netherlands}

\begin{abstract}
We formulate a comprehensive hydrodynamic theory of two-dimensional liquid crystals with generic $p-$fold rotational symmetry, also known as $p-$atics, of which mematics $(p=2)$ and hexatics $(p=6)$ are the two best known examples. Previous hydrodynamic theories of $p-$atics are characrerized by continuous ${\rm O}(2)$ rotational symmetry, which is higher than the discrete rotational symmetry of $p-$atic phases. By contrast, here we demonstrate that the discrete rotational symmetry allows the inclusion of additional terms in the hydrodynamic equations, which, in turn, lead to novel phenomena, such as the possibility of flow alignment at high shear rates, even for $p>2$. Furthermore, we show that any finite imposed shear will induce long-ranged orientational order in any $p-$atic liquid crystal, in contrast to the quasi-long-ranged order that occurs in the absence of shear. The induced order parameter scales like a non-universal power of the applied shear rate at small shear rates.
\end{abstract}

\maketitle

\section{Introduction}

The existence of the hexatic phase, i.e. a liquid-crystalline phase of two-dimensional matter intermediate between crystalline solid and isotropic liquid, was predicted by Halperin and Nelson in the late '70s \cite{Halperin:1978,Nelson:1979}, building upon Kosterlitz' and Thouless' groundbreaking discovery of defect-mediated phase-transitions in two dimensions \cite{Kosterlitz:1972,Kosterlitz:1973}, later refined by Young \cite{Young:1979}. According to this picture, known as KTHNY scenario, two-dimensional solids can melt via two distinct phase transitions as temperature is increased. First, the unbinding of neutral pairs and triplets of dislocations transforms a crystal, characterized by quasi-long-ranged translational order and long-ranged $6-$fold orientational order, into a hexatic liquid crystal, with quasi-long-ranged orientational order and short-ranged translational order. Second, as temperature is further increased, pairs of $5-$ and $7-$fold disclinations unbind, driving the transition of the hexatic liquid crystal into an isotropic liquid, in which both translational and orientational order are short-ranged. 

For the past four decades, the hexatic phase and KTHNY melting scenario have been subject to extensive theoretical and experimental investigation, aimed at clarifying the nature of the individual solid-hexatic and hexatic-isotropic phase transitions, as well as the role of  material properties. Large-scale numerical simulations \cite{Bladon:1995,Bernard:2011}, experiments with superparamagnetic colloids \cite{Zahn:1999, Gasser:2010} and, more recently, tilted monolayers of sedimented colloidal hard-spheres \cite{Thorneywork:2017}, in particular, have progressively shed light on several fascinating aspects of these transitions, while opening new avenues in condensed matter physics at the interface between statistical mechanics, material science and topology \cite{Anderson:2017,Bowick:2017,Beekman:2017,Sartori:2019,Maitra:2020, Mietke:2020}. By contrast, the hydrodynamic behavior of hexatics has received little attention and, with the exception of a small number of pioneering works, e.g. Refs.~\cite{Zippelius:1980a,Zippelius:1980b,Sonin:1998,Krieger:2014}, is  still  largely unexplored.

Yet, recent findings in tissue mechanics have renewed  interest in hexatic hydrodynamics, by providing this phase of matter with unexpected biological relevance. Like atomic, molecular and colloidal systems, that exhibit low temperature two-dimensional crystal phases, tissues are often neither ordered solids nor disordered liquids, but inhabit a continuum of intermediate states known as the epithelial-mesenchymal spectrum \cite{Zhang:2018}. This versatility lies at the heart of a myriad processes that are essential for life, such as embryonic morphogenesis \cite{Mongera:2018} and wound healing \cite{Brugues:2018}, as well as life-threatening conditions, such as metastatic cancer \cite{Zhang:2018}. Using a cell-resolved computational model of confluent tissues \cite{Nagai:2001,Farhadifar:2007}, Li and Pica Ciamarra have demonstrated that the solid and the isotropic liquid states of these model-epithelia are separated by an intermediate hexatic phase, in which cells are orientationally ordered and yet able to flow \cite{Li:2018}. Upon heating, the phase diagram is further enriched by various examples of phase coexistence, including solid-isotropic and hexatic-isotropic. This remarkable discovery sheds new light on the complex physics of tissues and, simultaneously provides a strong motivation for aiming at a deeper understanding of hexatic hydrodynamics and, more generally, of the hydrodynamics of liquid crystals with rotational symmetries other than polar (i.e. $1-$fold) and nematic (i.e. $2-$fold). 

\begin{figure}[b]
\centering
\includegraphics[width=\columnwidth]{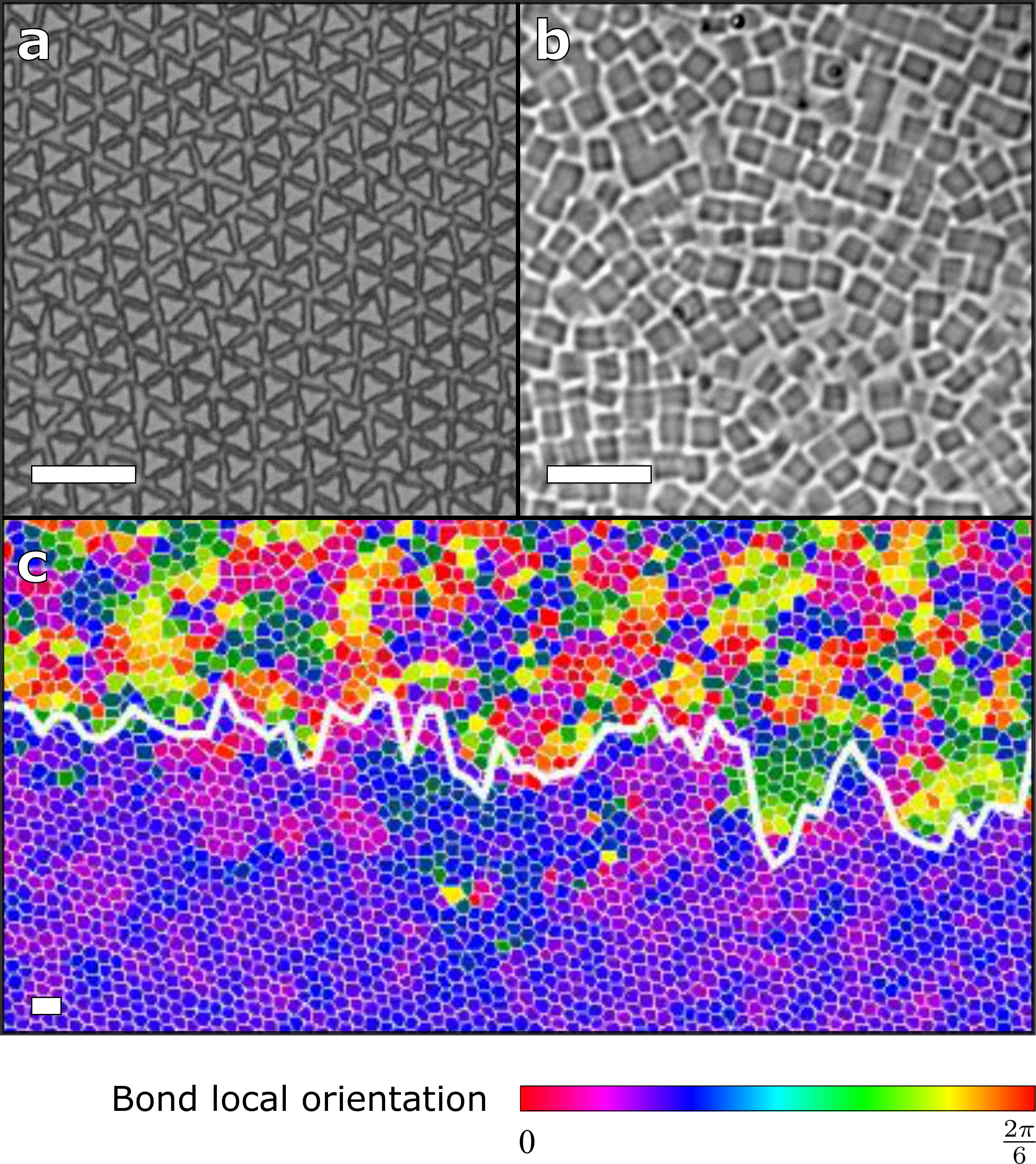}
\caption{\label{fig:1}Examples of $p-$atic colloidal suspensions. (a) Triatic ($p=3$) colloidal platelets (courtesy of Thomas Mason, adapted from Ref. \cite{Zhao:2012}). (b) Possible tetradic ($p=4$) suspension of colloidal cubes (courtesy of Janne-Mieke Meijer, adapted from Ref. \cite{Loffler:2018}). (c) Isotropic monolayer of sedimented colloidal hard spheres (top) coexisting with a hexatic phase ($p=6$, bottom) (courtesy of Roel Dullens, adapted from Ref. \cite{Thorneywork:2017}). In all panels, the scale bar corresponds to $10\,\mu$m.}	
\end{figure}

In this article we formulate a comprehensive hydrodynamic theory of two-dimensional liquid crystals endowed with $p-$fold rotational symmetry (i.e. symmetry with respect to rotations by $2\pi/p$), often referred to as $p-$atics (Fig. \ref{fig:1}). Previous hydrodynamic theories (e.g. Refs.~\cite{Zippelius:1980a,Zippelius:1980b,Sonin:1998,Krieger:2014}) are characterized by continuous ${\rm O}(2)$ rotational symmetry, which is higher symmetry than required by the discrete rotational symmetry of $p-$atics. By contrast, here we show that the discrete $p-$fold symmetry allows the inclusion of additional terms in the hydrodynamic equations, which, in turn, lead to novel phenomena, such as the possibility of flow alignment at high shear rates, even for arbitrary $p$ values (thus, in particular, for $p=6$). Our approach is based on a tensorial hydrodynamic variable, i.e. the $p-$atic tensor order parameter, that directly embodies the discrete rotational symmetry of $p-$atic phases. Exploiting the symmetries and the algebraic structure of this tensor, we will construct the equations governing the dynamics of $p-$atics, as well as the stress contribution associated with a departure from the $p-$atic ground state. In addition, we investigate the effects of an imposed shear flow on $p-$atic order. We restrict our attention exclusively to momentum-conserving systems. Experimental realizations of this would therefore have to be free standing films, to avoid loss of momentum due to friction with a substrate. Our two most striking conclusions can be summarized as follows. 

{\em 1)} An applied shear makes $p-$atic order long-ranged, in contrast to the quasi-long-ranged order that occurs in absence of shear. In particular, we find that at small shear rates $\deps$, the magnitude $\OP$ of the complex order parameter scales like a power law with the applied shear rate $\deps$: i.e. 
\begin{equation}\label{Psi(deps)intro}
\OP \sim \dot{\epsilon}^{\,\eta_{p}/4}\;,
\end{equation}
where $\eta_{p}<1/4$ is the non-universal, temperature-dependent exponent for the decay of orientational correlations in the absence of an applied shear \cite{Halperin:1978,Nelson:1979}. 

{\em 2)} In the presence of a simple shear flow, $p-$atics orient at specific angles with respect to the flow direction. This effect, referred to as ``flow alignment'' in the literature of nematic liquid crystals \cite{DeGennes:1993,Kleman:2003}, was only known for $p=1$ and $p=2$, where it occurs at arbitrary shear rate, unless the mesogens are anchored to a wall, which enforces a specific preferential direction. For $p>2$, on the other hand, flow alignment occurs exclusively if the the shear rate $\deps$ exceeds a given threshold.

The remainder of this paper is organized as follows. In Sec. \ref{sec:p-atic_tensor} we lay down our basic mathematical terminology and notation and introduce the $p-$atic order parameter tensor $\bm{Q}_{p}$, which plays a central role in our hydrodynamic theory. In Sec. \ref{sec:hydrodynamics}, we construct the hydrodynamic equations of $p-$atics and discuss about the additional terms arising when the $p-$fold rotation symmetry is fully taken into account. In Sec. \ref{sec:backflow} we investigate the effects of backflow, namely the hydrodynamic flow driven by spatial variations of $p-$atic order. In Sec. \ref{sec:long_range_order} we investigate the effect of flow on the orientational order of unconfined $p-$atics. Using fluctuating hydrodynamics and Renormalization Group (RG) arguments we demonstrate that, remarkably, a shear flow of arbitrary finite shear rate induces long-ranged orientational order in $p-$atics. In Sec. \ref{sec:flow_alignment} we consider two examples of viscous flow in $p-$atics, namely simple shear and Taylor-Couette flow, and demonstrate how flow alignment can arise at large shear-rates depending on the specific flow geometries and $p$ values. Finally, Sec. \ref{sec:conclusions} is devoted to conclusions. 

\section{\label{sec:p-atic_tensor}The $p-$atic tensor}

There are, at the moment, two approaches to describe structured fluids whose constituents have $p-$fold rotational symmetry that propagates over the macroscopic scale. The first approach relies on a tensor order parameters endowed with the same $p-$fold rotational symmetry. For instance, polar fluids (i.e. $p=1$) can be straightforwardly described in terms of a polarization vector, whereas nematic liquid crystals, which are invariant under $180^{\circ}$ rotations of the nematic director $\bm{n}$ (i.e. $p=2$), require a rank$-2$ traceless and symmetric tensor: i.e. $\bm{Q}_{2} = \OP (\bm{n}\otimes\bm{n}-\mathbb{1})$, with $\OP$ the scalar order parameter and $\mathbb{1}$ the identity tensor \cite{DeGennes:1993}. This tensor is invariant under the transformation $\bm{n}\rightarrow-\bm{n}$, thus it represents a suitable hydrodynamical variable to describe flow in nematics. The second approach, pioneered by Lammert {\em et al.} for nematics \cite{Lammert:1995}, and recently extended to describe phases characterized by generic three- dimensional point groups \cite{Liu:2016}, consists of a lattice-gauge formulation, in which a vectorial director field is coupled with auxiliary gauge fields, designed to implement the desired symmetry in the Hamiltonian. As the latter approach is inherently discrete, it cannot be integrated into a continuum mechanics framework. Therefore, we will adopt the former strategy and formulate a hydrodynamic theory using a tensor order parameter. 

\subsection{\label{sec:notation}Mathematical preliminaries and notation} 

In this Section we introduce the essential mathematical concepts and notation that will be used throughout the remainder of the article. The central object in our hydrodynamic theory of $p-$atic liquid crystals is a tensor order parameter endowed with the same $p-$fold rotational symmetry of the $p-$atic phase. In general, rank$-p$ tensors will be indicated as
\begin{equation}\label{eq:p_tensor}
\bm{T} = T_{i_{1}i_{2}\cdots\,i_{p}}\bm{e}_{i_{1}}\otimes\bm{e}_{i_{2}} \otimes\cdots\otimes\bm{e}_{i_{p}}\;,
\end{equation}
where $\bm{e}_{i_{n}}$, with $n=1,\,2\ldots\, p$, are basis vectors and summation over repeated indices is implied. Analogously, we define the $n-$th tensorial power of a generic tensor $\bm{T}$ as the $n-$fold product of the tensor with itself:
\begin{equation}
\bm{T}^{\otimes n} 
= \underbrace{\bm{T}\otimes\bm{T}\otimes\cdots\,\otimes\bm{T}}_{\text{$n$ times}}\;.
\end{equation}
Contracting one index of a generic rank$-p$ tensor, $\bm{T}$, with one index of a rank$-q$ tensor, $\bm{U}$, yields a rank$-(p+q-2)$ tensor. This operation will be indicated with a dot product, in analogy with vectorial and matrix multiplication. That is
\begin{equation}\label{eq:dot}
(\bm{T}\cdot \bm{U})_{i_{1}\cdots\,i_{p-1}j_{2}\cdots\,j_{q}} = T_{i_{1}\cdots\,i_{p-1}k}U_{kj_{2}\cdots\,j_{q}}\;.
\end{equation}
Similarly, the contraction of two indices will be indicated with
\begin{equation}\label{eq:colon}
(\bm{T} : \bm{U})_{i_{1}\cdots\,i_{p-2}j_{3}\cdots\,j_{q}} = T_{i_{1}\cdots\,i_{p-2}kl}U_{lkj_{3}\cdots\,j_{q}}\;.
\end{equation}
The inner product of two rank$-p$ tensors, on the other hand, will be denoted in the following  by the symbol $\odot$, that is
\begin{equation}\label{eq:odot}
\bm{T} \odot\bm{U} = T_{i_{1}i_{2}\cdots\,i_{p}}U_{i_{1}i_{2}\cdots\,i_{p}}\;.	
\end{equation}
In particular, the Euclidean norm of the tensor is given by
\begin{equation}\label{eq:norm}
\left|\bm{T}\right|^{2} = \bm{T}\odot\bm{T}\;.
\end{equation}	
Evidently, different choices of the dummy index $k$ and $l$ in Eqs. \eqref{eq:dot} and \eqref{eq:colon} yield different tensors, unless $\bm{T}$ and $\bm{U}$ are both symmetric. Although there is no unambiguous definition of a trace for rank$-p$ tensors with $p>2$, this exists for symmetric tensors, because of the symmetry under permutation of the indices. Consider then a symmetric rank$-p$ tensor $\bm{S}$, such that:
\begin{equation}
S_{i_{1}i_{2}\cdots\,i_{p}} = S_{i_{\sigma 1}i_{\sigma 2}\cdots\,i_{\sigma p}}\;, \qquad \sigma \in \mathfrak{S}_{p}\;,
\end{equation}
where $\mathfrak{S}_{p}$ is the group of permutations of $\{1,\,2\ldots\, p\}$. The trace of such a symmetric tensor is defined as the rank$-(p-2)$ tensor obtained upon contracting any two indices
\begin{equation}\label{eq:trace}
(\tr \bm{S})_{i_{1}i_{2}\cdots\,i_{p-2}} = S_{i_{1}i_{2}\cdots\,i_{p-2}jj}\;,
\end{equation}
and is symmetric by construction. 

Finally, we will denote with the symbol $\llbracket \cdots \rrbracket$ the operation of rendering an arbitrary rank$-p$ tensor symmetric and traceless. For $p\geq 2$ this can be achieved by contracting a rank$-p$ tensor $\bm{T}$ with the special rank$-2p$ tensor $\bm{\Delta}_{p,p}$, that is:
\begin{equation}\label{eq:traceless}
\traceless{T_{i_{1}i_{2}\cdots\,i_{p}}} = \Delta_{i_{1}i_{2}\cdots\,i_{p}j_{1}j_{2}\cdots\,j_{p}} T_{j_{1}j_{2}\cdots\,j_{p}}\;. 
\end{equation}
In three dimensions, an expression for $\bm{\Delta}_{p,p}$ was obtained in Ref. \cite{Hess:2015} using multipole potentials. An analogous expression can be obtained in two dimensions (see Appendix \ref{app:delta_tensor}):
\begin{equation}\label{eq:delta_tensor}
\bm{\Delta}_{p,p}
= \frac{(-1)^{p+1}}{p!(2p-2)!!}\,\nabla^{\otimes p}\left(r^{2p}\nabla^{\otimes p}\log\frac{r}{\ell}\right)\;,
\end{equation}
where $r=\sqrt{x^{2}+y^{2}}$ and $\ell$ an arbitrary length scale. $\bm{\Delta}_{p,p}$ is an {\em isotropic} tensor, that is a tensor whose structure is invariant upon rotation of the reference frame. It is symmetric with respect to any permutation of the first $p$ and the last $p$ indices: i.e. $i_{m} \leftrightarrow i_{n}$ and $j_{m} \leftrightarrow j_{n}$, with $m,n=1,\,2\ldots\, p$; as well as with respect to the exchange of the full set of $i-$ and $j-$indices: i.e. $\{i_{1}i_{2}\cdots\,i_{p}\}\leftrightarrow\{j_{1}j_{2}\cdots\,j_{p}\}$. Furthermore, contracting any pair among the first or last $p$ indices, yields the null tensor:
\begin{equation}
\Delta_{kki_{3}\cdots\,i_{p}j_{1}\cdots\,j_{p}} = \Delta_{i_{1}i_{2}\cdots\,i_{p}kkj_{3}\cdots\,j_{p}} = 0\;.
\end{equation}
This latter property, in particular, guarantees that:
\begin{equation}
\tr\traceless{\bm{T}} = \mathbb{0}_{p-2}\;,	
\end{equation}
where $\mathbb{0}_{p-2}$ is the rank$-(p-2)$ tensor whose elements are identically zero: i.e. $(\mathbb{0}_{p-2})_{i_{1}i_{2}\cdots i_{p-2}}=0$. For $p=1$, Eq. \eqref{eq:delta_tensor} yields the identity tensor: i.e. $\bm{\Delta}_{1,1}=\mathbb{1}$.

As noted by Park and Lubensky in Ref.~\cite{Park:1996}, various calculations involving traceless and symmetric rank$-p$ tensors can be conveniently performed by representing the tensor in terms of two circular basis vectors $\bm{\epsilon}_{\pm}$, defined as
\begin{equation}\label{circbasdef}
\bm{\epsilon}_{\pm}=\frac{\bm{e}_{x}\pm i\bm{e}_{y}}{\sqrt{2}}\,,
\end{equation}
which can readily be shown to satisfy the relations
\begin{subequations}\label{epsortho}
\begin{align}
\bm{\epsilon}_{+}\cdot\bm{\epsilon}_{+} &= \bm{\epsilon}_{-}\cdot\bm{\epsilon}_{-}=0\;,\\
\bm{\epsilon}_{+}\cdot\bm{\epsilon}_{-} &= 1\;,
\end{align}
\end{subequations}
as well as the tensorial identity
\begin{equation}\label{eps identity}
\bm{\epsilon}_{\pm}\otimes\bm{\epsilon}_{\mp} = \frac{\mathbb{1} \mp i\bm{\varepsilon}}{2}\;,
\end{equation}
with $\bm{\varepsilon}$ is the antisymmetric tensor: i.e. $\varepsilon_{xx}=\varepsilon_{yy}=0$ and $\varepsilon_{xy}=-\varepsilon_{yx}=1$. By virtue of Eq. \eqref{epsortho}, one can readily see that any rank$-p$ tensor of the form $\bm{T}=T{\bm \epsilon}_{\pm}^{\otimes p}$, with $T$ a scalar, is symmetric and traceless \cite{Zheng:1993}. Furthermore, it is possible to show that
\begin{equation}\label{eq:delta_circular}
\bm{\Delta}_{p,p}=\bm{\epsilon}_{-}^{\otimes p}\otimes\bm{\epsilon}_{+}^{\otimes p}+\bm{\epsilon}_{+}^{\otimes p}\otimes\bm{\epsilon}_{-}^{\otimes p}\;.
\end{equation}

\subsection{\label{sec:order_parameter}Order parameter tensor in $p-$atics} 

Let us consider a $p-$atic phase, whose microscopic constituents can be assigned a direction
\begin{equation}
\bm{\nu} = \cos\vartheta\,\bm{e}_{x}+\sin\vartheta\,\bm{e}_{y}\;.
\end{equation}
The latter may correspond to a particular direction at the molecular scale (e.g. the position of a specific functional group with respect to the center of mass of the molecule) or be conventionally assigned for perfectly $p-$fold symmetric constituents. Local $p-$atic order can then be identified starting from the following microscopic complex function \cite{Halperin:1978}:
\begin{equation}\label{eq:complex_order_parameter}
\psi_{p} = e^{ip\vartheta}\;.
\end{equation}
In two-dimensional equilibrium systems, $p-$atic order is quasi-long-ranged and is characterized by a power-law decaying two-point correlation function \cite{Halperin:1978,Nelson:1979}: 
\begin{equation}\label{eq:correlation_function}
\left\langle \psi_{p}^{*}(\bm{r})\psi_{p}(\bm{0}) \right\rangle \sim |\bm{r}|^{-\eta_{p}}\;,
\end{equation}
where $\langle \cdots \rangle$ denotes the ensemble average and $\eta_{p}$ is a positive non-universal exponent depending upon temperature and the $p-$atic orientational stiffness $K$ (see Sec. \ref{sec:hydrodynamic_equations}), given by
\begin{equation}\label{eq:eta}
\eta_{p}=\frac{p^{2}k_{B}T}{2\pi K} < \frac{1}{4}\;,	
\end{equation}
where the inequality prevents topological defects from unbinding at equilibrium \cite{Kosterlitz:1974}. The $p-$atic order parameter can be expressed as:
\begin{equation}\label{eq:complex_order_parameter}
\Psi_{p} = \langle \psi_{p} \rangle = \OP e^{i p \theta}\;,
\end{equation}
where $\OP$ and $\theta$ are, respectively, the scalar order parameter amplitude and average orientation. Quasi-long-range order implies that $\langle \psi_{p} \rangle$ is scale-dependent and vanishes at large length scales (see Sec. \ref{sec:long_range_order} and Ref. \cite{Udink:1987}). Specifically,
\begin{equation}\label{eq:order_parameter_scale}
\OP \sim \left(\frac{a}{\ell}\right)^{\eta_{p}/2}\;,
\end{equation}
where $a$ is a short distance (i.e. ultraviolet) cut-off and $\ell$ the length scale at which the system is probed.

\begin{figure}[t]
\centering
\includegraphics[width=\columnwidth]{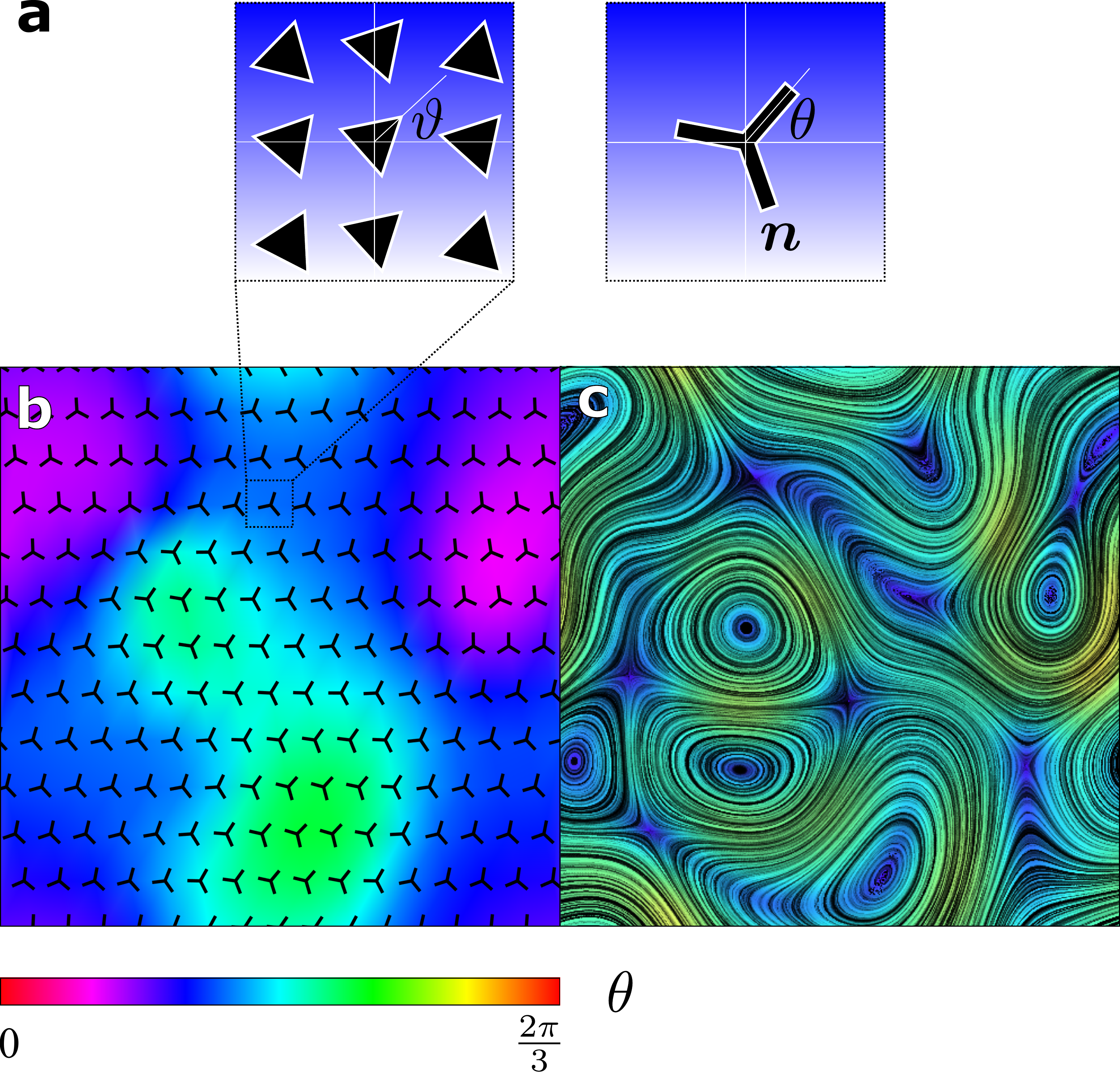}
\caption{\label{fig:2}(a) Schematic illustration of triatic building blocks (left) together with the corresponding coarse-grained $p-$atic director (right). The molecular and average orientations, here denoted as $\vartheta$ and $\theta$ respectively, are related by Eq. \eqref{eq:complex_order_parameter}. (b,c) Typical configuration of the triatic director (b) and velocity field (c) coarsening from an initially disordered state. The data in displayed in panels (b) and (c) have been obtain by a numerical integration of the hydrodynamic equation given in Sec. \ref{sec:hydrodynamics}.}	
\end{figure}

As in nematics, a symmetric and traceless $p-$atic order parameter tensor can be constructed by averaging the $p-$th degree tensorial powers of the microscopic direction $\bm{\nu}$ within a fluid element, this being defined as a portion of the system that is sufficiently small to be considered infinitesimal with respect to the system size and yet sufficiently large to contain a macroscopic number of $p-$atic building blocks:
\begin{equation}\label{eq:p-atic_tensor_1}
\bm{Q}_{p} 
= \sqrt{2^{p-2}}\,\traceless{\langle \bm{\nu}^{\otimes p} \rangle}\;
= \sqrt{2^{p-2}}\,\OP\traceless{\bm{n}^{\otimes p}}\;,
\end{equation}
where 
\begin{equation}
\bm{n} = \cos\theta\,\bm{e}_{x}+\sin\theta\,\bm{e}_{y}
\end{equation}
is the $p-$atic director field (Fig. \ref{fig:2}a). Consistently with the standard convention in nematics, the numerical pre-factor has been chosen in such a way as to obtain:
\begin{equation}\label{eq:qp-psi}
|\bm{Q}_{p}|^{2} = \frac{\OP^{2}}{2}\;.
\end{equation}
Furthermore, using the auto-orthogonality of circular basis vectors, embodied in Eqs. \eqref{epsortho}, one can express the $p-$atic order parameter tensor in the following equivalent forms:
\begin{align}
\label{eq:p-atic_tensor_2}
\bm{Q}_{p} 
&= \Re \left[\Psi_{p}\bm{\epsilon}_{-}^{\otimes p} \right] \notag \\[5pt]
&= \Re \left[\Psi_{p}^{*}\bm{\epsilon}_{+}^{\otimes p} \right]\; \notag \\
&= \frac{1}{2}\left(\Psi_{p}\bm{\epsilon}_{-}^{\otimes p}+\Psi^*_{p}\bm{\epsilon}_{+}^{\otimes p} \right)\;,
\end{align}
where $\Re[\cdots]$ yields real part of any complex quantity, whereas $\Im[\cdots]$ will be used for the imaginary part. For $\OP=\sqrt{2}$, these expressions coincide the $p-$atic tensor introduced in Ref.~\cite{Park:1996}. Finally, using standard algebraic manipulations, one can prove that contracting the $\bm{Q}_{p}$ tensor with itself yields the isotropic tensor, that is:
\begin{equation}\label{eq:q_dot_q}
\bm{Q}_{p}\cdot\bm{Q}_{p} = \frac{\OP^{2}}{4}\bm{\Delta}_{p-1,p-1}\;.	
\end{equation}
As we will see in the next Section, Eq. \eqref{eq:q_dot_q} has implications for the structure of the viscosity tensor and other coupling tensors of the theory.
 
\section{\label{sec:hydrodynamics}Hydrodynamic equations of two-dimensional $p-$atics}

\subsection{\label{sec:hydrodynamic_variables}Hydrodynamic variables}

Our goal  is to describe the spatiotemporal evolution of $p-$atic liquid crystals at large length and long time scales. To this end, we start by identifying a set of ``hydrodynamic variables", namely material fields whose evolution rate vanishes as the length scale at which they are probed diverges (see e.g. Ref. \cite{Forster:1975}). Whereas almost any variable determined by a macroscopic number of degrees of freedom relaxes to its equilibrium value on microscopic time scales, these ``slow'' variables naturally arise in critical phenomena (see e.g. Ref. \cite{Chaikin:1995}) or, away from criticality, in the presence of conservation laws and broken continuous symmetries. In the absence of external stimuli, films of $p-$atic liquid crystals are characterized by four conserved quantities, namely the system's total mass $M=\int {\rm d}A\,\rho$, momentum $\bm{\mathcal{P}}=\int {\rm d}^{2}r\,\rho\bm{v}$, energy $E=\int{\rm d}^{2}r\,\rho e$ and entropy $S=\int {\rm d}^{2}r\,\rho s$, as well as a broken rotational symmetry, embodied in the $p-$atic tensor $\bm{Q}_{p}$ or, equivalently, in the complex order parameter $\Psi_{p}$, defined in Eq. \eqref{eq:complex_order_parameter}.

The hydrodynamic equations governing the evolution of the density fields associated with conserved quantities, namely $\rho$, $\rho\bm{v}$, $e$ and $s$, follow directly from the fundamental laws of continuum mechanics (see e.g. Ref. \cite{Gallavotti:2002}) and are given by:
\begin{subequations}\label{eq:generic_hydrodynamics}
\begin{gather}
\frac{D\rho}{Dt} + \rho \nabla\cdot\bm{v}= 0\;,\\
\rho\frac{D\bm{v}}{Dt} = \nabla\cdot\bm{\sigma}+\bm{f}\;,\\
\rho \frac{De}{Dt} + \nabla\cdot\bm{\mathcal{Q}} = \bm{\sigma}:\nabla\bm{v}\;,\\
\rho T \frac{Ds}{Dt} + \nabla\cdot\bm{\mathcal{Q}} = \bm{\sigma}^{({\rm v})}:\nabla\bm{v}+2R\;,
\end{gather}
\end{subequations}
where $D/Dt=\partial_{t}+\bm{v}\cdot\nabla$ is the material derivative, $\bm{\sigma}$ the stress tensor, $\bm{f}$ the external force per unit area, $\bm{\mathcal{Q}}$ the heat flux density, resulting from local energy and entropy variations, and $2R\ge 0$ the entropy production rate. The stress tensor is customarily decomposed into a {\em reactive} component $\bm{\sigma}^{({\rm r})}$, resulting from the reversible processes, and a {\em viscous} component $\bm{\sigma}^{({\rm v})}$, arising from the irreversible processes, which give rise to local entropy production. Thus 
\begin{equation}\label{eq:stress_tensor}
\bm{\sigma} = \bm{\sigma}^{({\rm r})}+\bm{\sigma}^{({\rm v})}\;.
\end{equation}
Eqs. \eqref{eq:generic_hydrodynamics} must be complemented with the equation governing the dynamics of the $p-$atic tensor $\bm{Q}_{p}$ and a constitutive equation for the stress tensor $\bm{\sigma}$ in terms of the other hydrodynamic variables. Both tasks will be accomplished in the next Subsection. 

\subsection{\label{sec:hydrodynamic_equations}Hydrodynamics of the $p-$atic tensor}

The hydrodynamic equation describing the spatiotemporal evolution of the broken symmetry variable $\Psi_{p}$ are most conveniently derived in terms of the order parameter tensor $\bm{Q}_{p}$, defined in Sec. \ref{sec:order_parameter}, by taking advantage of the algebraic structure of the tensor in order to achieve frame invariance. Specifically, as $\bm{Q}_{p}$ is traceless and symmetric, one can construct its hydrodynamic equation by expressing its time derivative as a sum of all possible symmetric and traceless rank$-p$ tensor combinations of the velocity gradient tensors $\nabla\bm{v}$ and $\bm{Q}_{p}$ and its gradients. In nematics, this procedure, as explained in, e.g., Ref. \cite{Olmsted:1992}, leads to the following well established parabolic partial differential equation:
\begin{equation}\label{eq:nematodynamics}
\frac{{\rm d}\bm{Q}_{2}}{{\rm d}t} = \Gamma\bm{H}_{2}+\lambda_{2} \traceless{\boldsymbol{u}}+\bar{\lambda}_{2}\tr(\bm{u})\bm{Q}_{2}\;,
\end{equation}
where the left-hand side indicates the corotational time derivative of a rank$-2$ tensor field
\begin{equation}
\frac{{\rm d}\bm{Q}_{2}}{{\rm d}t} = \frac{D\bm{Q}_{2}}{Dt}-\bm{Q}_{2}\cdot\bm{\omega}+\bm{\omega}\cdot\bm{Q}_{2}\;,
\end{equation} 
where $\bm{\omega}=[\nabla\bm{v}-(\nabla\bm{v})^{\rm T}]/2$, with ${\rm T}$ denoting transposition, is the vorticity tensor. On the right-hand side of Eq. \eqref{eq:nematodynamics}, $\bm{H}_{2}=-\delta F/\delta\bm{Q}_{2}$ is the molecular tensor describing the relaxation of the nematic phase toward the minimum of the free energy $F$, with $\Gamma^{-1}$ a rotational viscosity, $\bm{u}=[\nabla\bm{v}+(\nabla\bm{v})^{\rm T}]/2$ is the strain-rate tensor and $\lambda_{2}$ and $\bar{\lambda}_{2}$ are dimensionless constants. The quantity $\lambda_{2}$, in particular, is referred to as the flow alignment parameter and it can cause the nematic director to align with an imposed shear flow (see e.g. Ref. \cite{DeGennes:1993}, and Sec. \ref{sec:flow_alignment}). The last term on the right-hand side of Eq. \eqref{eq:nematodynamics}, on the other hand, affects the magnitude of the scalar order parameter and vanishes identically in the case of incompressible flow, where $\tr(\bm{u})=\nabla\cdot\bm{v}=0$. Notably, the first term on the right-hand side of Eq. \eqref{eq:nematodynamics} has the opposite signature with respect to ${\rm d}\bm{Q}_{2}/{\rm d}t$ under time reversal, whereas the last two terms have the same signature. Hence, these terms  embody irreversible (dissipative) and reversible (reactive) processes respectively.

A hydrodynamic equation such as Eq. \eqref{eq:nematodynamics} could, in principle, be formulated for any $p-$atic liquid crystal upon constructing all possible rank$-p$ tensors, obtained by contracting $\nabla\bm{v}$ and $\bm{Q}_{p}$, that are simultaneously symmetric and traceless. In the following, we demonstrate that such a hydrodynamic equation can indeed be constructed in the form
\begin{equation}
\label{eq:p-atics_hydrodynamics}
\frac{{\rm d}\bm{Q}_{p}}{{\rm d}t} = \Gamma\bm{H}_{p} + \bm{L}_{p}+\bm{N}_{p}\;,
\end{equation}
where $\bm{H}_{p}=-\delta F/\delta \bm{Q}_{p}$ is a $p-$atic generalization of the molecular tensor and $\bm{L}_{p}$ and $\bm{N}_{p}$ are, respectively, linear and nonlinear tensorial functions of the strain-rate $\bm{u}$. The Landau free energy $F=\int{\rm d}^{2}r\,f$ can be readily constructed from the free energy density
\begin{align}\label{eq:free_energy}
f 
&= \frac{1}{2}\,L\left| \nabla\bm{Q}_{p} \right|^{2}+\frac{1}{2}\,a_{2}\left|\bm{Q}_{p} \right|^{2} + \frac{1}{4}\,a_{4}\left|\bm{Q}_{p} \right|^{4} \notag \\
&= \frac{1}{4}\,L\left| \nabla\Psi_{p} \right|^{2}+\frac{1}{4}\,a_{2}\left|\Psi_{p} \right|^{2} + \frac{1}{16}\,a_{4}\left|\Psi_{p} \right|^{4}\;,
\end{align}
where have made use of Eq. \eqref{eq:qp-psi} to derive the second equality. The constant $L$ is the order parameter stiffness of $p-$atic phases, while the phenomenological coefficients $a_{2}$ and $a_{4}$ favor a non-vanishing $\OP$ value in the ordered phase (i.e. where $a_{2}<0$), away from the system boundary or topological defects. Specifically
\begin{equation}\label{eq:bare_order_parameter}
\OPeq = \sqrt{-\frac{2a_{2}}{a_{4}}}\;,
\end{equation}
at the minumum of the free energy. We stress that $\OPeq$ is the order parameter magnitude at the scale of the ultraviolet cut-off $a$, introduced in Eq. \eqref{eq:order_parameter_scale}, and should not be confused with the renormalized order parameter $\OP$, which, as explained in Sec. \eqref{sec:order_parameter}, vanishes in the thermodynamic limit. Cubic terms, such as those obtained upon contracting the tensor $\bm{Q}_{p}^{\otimes 3}$, cannot be constructed for odd $p$ values and, using Eqs. \eqref{epsortho} and \eqref{eq:p-atic_tensor_2}, can be shown to vanish identically for even $p$ values in two dimensions. In the case of two-dimensional nematics (i.e. $p=2$), the free energy density, Eq. \eqref{eq:free_energy}, can be augmented with additional elastic terms, such as $Q_{ij}\partial_{i}Q_{kl}\partial_{j}Q_{kl}$, to independently account for the costs of {\em bending} (i.e. longitudinal) and {\em splay} (i.e. transverse) deformations \cite{Schiele:1983}. For $p>2$, such a construction is not possible and the system is elastically isotropic, consistently with the intuition that a notion of longitudinal and transverse directions can be unambiguously defined only for rod-shaped objects.

From the free energy Eq. \eqref{eq:free_energy}, one obtains
\begin{equation}\label{eq:molecular_tensor}
\bm{H}_{p} = L \nabla^{2}\bm{Q}_{p}-(a_{2}+a_{4}|\bm{Q}_{p}|^{2})\bm{Q}_{p}\;,
\end{equation}	
which is symmetric and traceless because $\bm{Q}_{p}$ is. Similarly, a $p-$atic generalization of the corotational derivative can be constructed starting from the generic expression
\begin{equation}\label{eq:corotational_derivative}
\frac{{\rm d}\bm{Q}_{p}}{{\rm d}t} = \frac{D\bm{Q}_{p}}{Dt} - \kappa \traceless{\bm{Q}_{p}\cdot\bm{\omega}}\;,
\end{equation}
where $\kappa$ is a numerical pre-factor, which can be determined as follows. Consider a system in which $\OP$ is uniform throughout the system. Then, using Eqs. \eqref{eq:p-atic_tensor_2} and \eqref{eq:corotational_derivative}, contracting both sides of the resultant equation with $\bm{\epsilon}_{+}^{\otimes p}$ and using the orthogonality relation Eqs. \eqref{epsortho}, one can cast Eq. \eqref{eq:corotational_derivative} in the form:
\begin{equation}\label{eq:dthetadt1}
\frac{{\rm d}\theta}{{\rm d}t} = \frac{D\theta}{Dt}-\frac{\kappa}{p}\,\omega_{xy}\;. 
\end{equation}
The last term on the right-hand side of this equation describes the effect of rigid-body rotations on the $p-$atic director. It must be equal to $\omega_{xy}$, hence $\kappa=p$. 

Now, taking this into account and momentarily ignoring the tensors $\bm{L}_{p}$ and $\bm{N}_{p}$ in Eq. \eqref{eq:p-atics_hydrodynamics} yields the following equation for the local average orientation $\theta$:
\begin{equation}\label{eq:dthetadt2}
\frac{D\theta}{Dt} = \frac{K}{\gamma}\,\nabla^{2}\theta+\omega_{xy}\;,
\end{equation}
where the orientational stiffness $K$ [see Eq. \eqref{eq:eta}] and the rotational viscosity $\gamma$ are given by:
\begin{equation}\label{eq:material_parameters}
K = \frac{p^{2}\OP^{2}}{2}\,L\;,\qquad
\gamma = \frac{p^{2}\OP^{2}}{2}\,\Gamma^{-1}\;.
\end{equation}
For $p=6$, Eq. \eqref{eq:dthetadt2} coincides with the hydrodynamic equation for hexatics first proposed by Zippelius {\em et al}. \cite{Zippelius:1980a} and later adopted in Refs. \cite{Zippelius:1980b,Sonin:1998,Krieger:2014}. In the following, we will demonstrate that Eq. \eqref{eq:dthetadt2} can in fact be augmented by additional terms, originating from the interplay between orientational order and flow and embodied in the tensors $\bm{L}_{p}$ and $\bm{N}_{p}$. For $p>3$, these terms depend upon high order derivatives of the velocity field, or nonlinear powers of the strain rate and, unlike in nematics, are ``irrelevant'' in the RG sense of not altering the scaling or form of equilibrium correlation functions at large length and long time scales. However, in driven systems, in particular those subject to externally imposed shear flows, they can lead to new phenomena, including flow alignment, in contrast to the predictions of the linear theory \cite{Zippelius:1980a}. 

Although unknown a priori, the tensors $\bm{L}_{p}$ and $\bm{N}_{p}$ in Eq. \eqref{eq:p-atics_hydrodynamics} can be expanded in gradients of the velocity field or, analogously, of powers of the strain rate tensor $\bm{u}$. Each gradient term is proportional to the wave-number $|\bm{q}|=2\pi/\ell$, with $\ell$ the length scale under consideration, in the Fourier expansion of the velocity. Therefore, at large length scales, one can truncate the expansion at the lowest-order term whose symmetric and traceless part is non-zero. For $p=2$, for instance, the coupling can be expressed, at the lowest order in both $\bm{q}$ and $\bm{Q}_{2}$ in terms of the isotropic $\bm{\Delta}_{p,p}$ tensor introduced in Sec. \ref{sec:notation}.
\begin{align}\label{eq:lambda_2}
\bm{L}_{2} 
&= \left(\lambda_{2} \bm{\Delta}_{2,2}+\bar{\lambda}_{2}\bm{Q}_{2}\otimes\mathbb{1}\right):\bm{u} \notag\\[5pt] 
&= \lambda_{2}\traceless{\bm{u}}+\bar{\lambda}_{2}\tr(\bm{u})\bm{Q}_{2}+\mathcal{O}\left(|\bm{q}|^{2}\right)\;,
\end{align}
consistent with Eq. \eqref{eq:nematodynamics}. We stress that the isotropy of the $\bm{\Delta}_{2,2}$ tensor guarantees that the resulting hydrodynamic equation is frame invariant and is thus indispensable in this construction. Now, although higher order isotropic tensors can be obtained from $\bm{\Delta}_{p,p}$, with the exception for the $p=2$ case, this yields tensors whose rank $2p$ is higher than the value $p+2$ required to couple $\bm{Q}_{p}$ and $\bm{u}$. For even $p$ values, $\bm{\Delta}_{p,p}$ could be contracted with another isotropic tensor of rank$-(p-2)$, but the only one available in two dimensions is $\bm{\Delta}_{p/2-1,p/2-1}$ and the tensor resulting from this contraction is either null or anisotropic (i.e. not frame invariant). For odd $p$ values, no isotropic tensor exists such that, when contracted with $\bm{\Delta}_{p,p}$, yields a rank$-(p+2)$ tensor. From this we conclude that an $\mathcal{O}(|\bm{q}|)$ coupling between $p-$atic order and flow, such as that given by Eq. \eqref{eq:lambda_2}, does not exist for any $p>2-$atic liquid crystal. 

In contrast, various flow alignment terms can be constructed of the form:
\begin{equation}\label{eq:flow_alignment}
\bigotimes_{i=1}^{\infty}\traceless{ \nabla^{\otimes\alpha_{i}}\bm{u}^{\otimes \beta_{i}}}
= \mathcal{O}\left(|\bm{q}|^{\alpha_{1}+\beta_{1}+\alpha_{2}+\beta_{2}\cdots}\right)\;,
\end{equation}
where the exponents $(\alpha_{i}, \beta_{i})\in\mathbb{N}$ are solutions of the Diophantine equation
\begin{equation}\label{eq:diophantine}
\sum_{i}(\alpha_{i}+2\beta_{i})(1-\delta_{0,\beta_{i}}) = p\;,
\end{equation}
and we use the convention $(\cdots)^{\otimes 0}=1$. Now, the only term of this form linear in $\bm{u}$ is obtained when $\beta_{1}=1$ and $\alpha_{1}=p-2$. Thus, the linear flow alignment tensor $\bm{L}_{p}$ is given by: 
\begin{equation}\label{eq:l_tensor}
\bm{L}_{p} = \lambda_{p}  \big\llbracket \nabla^{\otimes p-2}\bm{u}  \big\rrbracket + \bar{\lambda}_{p}\tr(\bm{u})\bm{Q}_{p}\;,
\end{equation}
with $\lambda_{p}$ and $\bar{\lambda}_{p}$ phenomenological constants. As we will shortly demonstrate, this tensor represents the only reactive coupling between orientational order and flow that all $p-$atics have in common. Notice that $\traceless{\bm{Q}_{p}\cdot\bm{u}}=1/2\,\tr(\bm{u})\bm{Q}_{p}$, as one can demonstrate using the representations Eqs. \eqref{eq:p-atic_tensor_2} and \eqref{eq:delta_circular} of the tensor $\bm{\Delta}_{p,p}$ and the order parameter tensor, respectively, in terms of the circular basis vectors, as well as Eqs. \eqref{epsortho} and \eqref{eps identity}. Thus, the linear flow alignment tensor $\bm{L}_{p}$ has no contribution other than those featured in Eq. \eqref{eq:l_tensor}.

Conversely, for sufficiently large $p$ values, Eqs. \eqref{eq:flow_alignment} and \eqref{eq:diophantine} give rise to several nonlinear terms that, unlike Eq. \eqref{eq:l_tensor}, differ depending on whether $p$ is even or odd and correspond to both reversible and irreversible processes. At the lowest order in $\bm{q}$, these can be expressed in the generic form:
\begin{equation}\label{eq:n_tensor}
\bm{N}_{p} = \nu_{p}  \big\llbracket \nabla^{\otimes p\;{\rm mod}\;2}\bm{u}^{\lfloor p/2 \rfloor} \big\rrbracket \;,	
\end{equation}	
where $\nu_{p}$ is another phenomenological constant, $\lfloor \cdot \rfloor$ denotes the floor function and $p\;{\rm mod}\;2 = p-2 \lfloor p/2\rfloor$ is zero for even $p$ values and one for odd $p$ values. 

Some examples are in order. For $p=2$ and $3$, the only solutions of Eq. \eqref{eq:diophantine} are, respectively, $(\alpha_{1},\beta_{1})=(0,1)$ and $(\alpha_{1},\beta_{1})=(1,1)$. Thus, the coupling between local orientation and flow is, at this order, embodied solely in the linear flow alignment tensor, Eq. \eqref{eq:l_tensor}, whereas $\nu_{2}=0$ and $\nu_{3}=0$. For $p=4$, on the other hand, Eq. \eqref{eq:diophantine} has two independent solutions, $(\alpha_{1},\beta_{1})=(2,1)$ and $(0,2)$ and the corresponding nonlinear flow alignment tensor is given, at the lowest order in $\bm{q}$, by
\begin{equation}\label{eq:n_tensor_4}
\bm{N}_{4} = \nu_{4}\traceless{\bm{u}^{\otimes 2}}\;.
\end{equation}
As this tensor has opposite signature with respect to ${\rm d}\bm{Q}_{4}/{\rm d}t$ under time-reversal, it describes an irreversible process originating from the interplay between tetradic order and flow, with no counterpart in either nematics or triatics. For $p=5$, Eq. \eqref{eq:diophantine} has instead three independent solutions: $(\alpha_{1},\beta_{1},\alpha_{2},\beta_{2})=(3,1,0,0)$, $(1,2,0,0)$ and $(0,1,1,1)$. The last two of these are both of order $\mathcal{O}\left(|\bm{q}|^{3}\right)$ and feature, in general, different terms, but yield the same function of $\bm{u}$ under the action of the $\traceless{\cdots}$ operator. Thus:
\begin{equation}\label{eq:n_tensor_5}
\bm{N}_{5} =\nu_{5} \traceless{\nabla\bm{u}^{\otimes 2}}\;.
\end{equation}
Similarly, for the most physically relevant case $p=6$, Eq. \eqref{eq:diophantine} has five independent solutions: $(\alpha_{1},\beta_{1},\alpha_{2},\beta_{2})=(4,1,0,0)$, $(0,3,0,0)$, $(1,1,1,1)$, $(0,1,2,1)$ and $(2,2,0,0)$ from which, at the lowest order in $\bm{q}$, one finds:
\begin{equation}\label{eq:n_tensor_6}
\bm{N}_{6} 
= \nu_{6} \traceless{\bm{u}^{\otimes 3}}\;.
\end{equation}
Unlike the previous cases, $\bm{N}_{6}$ is odd under time reversal, thus it describes a reversible process. 

In summary, the dynamics of the $p-$atic tensor can generally be described by the following partial differential equation:
\begin{multline}\label{eq:p-atic_hydrodynamics}
\frac{D\bm{Q}_{p}}{D t} 
=  \Gamma\bm{H}_{p} + p \big \llbracket \bm{Q}_{p}\cdot\bm{\omega} \big \rrbracket 
+ \bar{\lambda}_{p}\tr(\bm{u})\bm{Q}_{p} \\
+ \lambda_{p} \big\llbracket \nabla^{\otimes p-2}\bm{u} \big\rrbracket 
+ \nu_{p} \big \llbracket \nabla^{\otimes p\,{\rm mod}\,2} \bm{u}^{\lfloor p/2 \rfloor} \big \rrbracket\;.
\end{multline} 
The apparent complexity of Eq. \eqref{eq:p-atic_hydrodynamics} simplifies considerably when expressed in terms of the complex order parameter $\Psi_{p}$, as given in Eqs. \eqref{eq:p-atic_tensor_1} and \eqref{eq:p-atic_tensor_2}. This can be done by inserting Eq. \eqref{eq:p-atic_tensor_2} into Eq. \eqref{eq:p-atic_hydrodynamics}, and then contracting both sides of the resultant equation with $\bm{\epsilon}_{+}^{\otimes p}$. Using the orthogonality relations, Eqs. \eqref{epsortho}, then yields
\begin{multline}\label{eq:psi}
\frac{D\Psi_{p}}{Dt} 
%= \Gamma L\nabla^{2}\Psi_{p}
= 2\Gamma L\partial\bar{\partial}\Psi_{p}
- \Gamma\left(a_{2}
+ \frac{a_{4}}{2}\,\OP^{2}\right)\Psi_{p} 
+ ip\,\omega_{xy}\Psi_{p} \\
+ \bar{\lambda}_{p}\tr(\bm{u})\Psi_{p}
+ 2\lambda_{p}\partial^{p-2}\U
+ 2\nu_{p}\partial^{\,p\,{\rm mod}\,2}\,\U^{\lfloor p/2 \rfloor}\;,
\end{multline}
where we have introduced the complex strain-rate:
\begin{equation}\label{eq:complex_strain_rate}
\U = \left(\bm{\epsilon}_{+}\otimes\bm{\epsilon}_{+}\right):\bm{u} = \frac{u_{xx}-u_{yy}}{2}+iu_{xy}\;,
\end{equation}
as well as the complex derivative $\partial=(\partial_{x}+i\partial_{y})/\sqrt{2}$ and its conjugate $\bar{\partial}=(\partial_{x}-i\partial_{y})/2$.

For $p=2$, Eq. \eqref{eq:p-atic_hydrodynamics} reduces to the classic hydrodynamic equation for the nematic tensor, Eq. \eqref{eq:nematodynamics}. For $p > 2$, on the other hand, Eqs. \eqref{eq:p-atic_hydrodynamics} and \eqref{eq:psi} provide a generalization of Eq. \eqref{eq:dthetadt2}, in which the interplay between $p-$atic and flow is not limited to the precession of the local orientation $\theta$ in the vorticity field, but includes couplings with the local strain rate, whose strength is set by the material parameters $\lambda_{p}$, $\bar{\lambda}_{p}$ and $\nu_{p}$. With exception of $\lambda_{2}$ and $\bar{\lambda}_{p}$, which are dimensionless numbers, the parameters $\lambda_{p}$ and $\nu_{p}$ depend upon intrinsic length and time scales. Denoting these with $\ell$ and $\tau$, one has
\begin{equation}
\lambda_{p} \sim \ell^{p-2}\;,\qquad
\nu_{p} \sim \ell^{p\,{\rm mod}\,2}\,\tau^{\lfloor p/2 \rfloor-1}\;. 
\end{equation}
As a consequence, the linear flow alignment terms become relevant when the strain rate $\bm{u}$ undergoes spatial variations over a length scale of order $\ell$, whereas the nonlinear terms yield measurable effects when $\bm{u}$ is comparable in magnitude with $1/\tau$: i.e. $\dot{\epsilon}\tau\approx 1$, where $\deps$ is the typical magnitude of $u_{xx}$, $u_{yy}$ and $u_{xy}$. Thus, unlike in nematics, hydrodynamic flow may or may not affect the dynamics of the $p-$atic director in ways other than the simple advection and precession, depending on the specific value of the material parameters. In some, but not all, cases, these couplings between flow and orientation may never lead to measurable effects. Furthermore, $\lambda_{p}$ and $\nu_{p}$ could, in principle, depend on the shear rate $\dot{\epsilon}$, e.g. 
\begin{subequations}\label{eq:lambda_nu_effective}
\begin{gather}
\lambda_{p}=\lambda_{p}^{(0)}+\lambda_{p}^{(1)}\deps+\lambda_{p}^{(2)}\deps^{2}+\cdots\;,\\[5pt]
\nu_{p}=\nu_{p}^{(0)}+\nu_{p}^{(1)}\deps+\nu_{p}^{(2)}\deps^{2}+\cdots\;.
\end{gather}
\end{subequations}
For small shear rates, higher order terms are evidently unimportant, but the same argument does not apply at large shear rates since these terms would become comparable in magnitude to those in Eqs. \eqref{eq:p-atic_hydrodynamics} and \eqref{eq:psi}. As we will discuss in Sec. \ref{sec:flow_alignment}, these considerations are particularly important in the context of flow alignment. 

Finally, the case $p=1$ is sufficiently different from the rest to justify a separate treatment. In this case, the order parameter is the polarization vector $\bm{P}=\OP\bm{n}$, whose hydrodynamic equation can be obtained, on the basis of phenomenological arguments (e.g. Ref. \cite{Kruse:2004}) or microscopic models (e.g. Ref. \cite{Kung:2006}), as follows
\begin{equation}\label{eq:p}
\frac{D\bm{P}}{Dt} = \Gamma\bm{H}_{1}-\bm{\omega}\cdot\bm{P}+\lambda_{1} \bm{u}\cdot\bm{P}+\bar{\lambda}_{1}\tr(\bm{u})\bm{P}\;,
\end{equation}
where $\bm{H}_{1}=-\delta F/\delta\bm{P}$ is the molecular field. Under the assumption of equal splay and bending moduli, the free-energy density can be expressed as
\begin{equation}
f = \frac{1}{2}\,L\,|\nabla\bm{P}|^{2}+\frac{1}{2}\,a_{2}|\bm{P}|^{2}+\frac{1}{4}\,a_{4}|\bm{P}|^{4}\;,
\end{equation}
from which $\bm{H}_{1}$ can be readily found in the form
\begin{equation}
\bm{H}_{1} = L\nabla^{2}\bm{P}\green{-}(a_{2}+a_{4}|\bm{P}|^{2})\bm{P}\;.
\end{equation}
This finally allows us to cast Eq. \eqref{eq:p} in terms of the polar complex order parameter $\Psi_{1}=|\Psi|\exp i\theta$:
\begin{multline}\label{eq:psi_1}
\frac{D\Psi_{1}}{Dt}
= 2\Gamma L\partial\bar{\partial}\Psi_{1}
- \Gamma\left(a_{2}
+ a_{4}\OP^{2}\right)\Psi_{1} \\
+ i\,\omega_{xy}\Psi_{1}
+ \lambda_{1}\U\Psi_{1}^{*}+\left(\frac{\lambda_{1}}{2}
+ \bar{\lambda}_{1}\right)\tr(\bm{u})\Psi_{1}\;.
\end{multline}
We stress that the cases $p=1,\,2$ are the only ones for which the complex strain rate $\U$ is linearly coupled to the order parameter $\psi_{p}$ at leading order in derivatives. As we will see in Secs. \ref{sec:long_range_order} and \ref{sec:flow_alignment}, this peculiarity of polar and nematic liquid crystals crucially affects the onset of flow alignment.
 
\subsection{\label{sec:theta}Hydrodynamic equations for the orientation field}

Eqs. \eqref{eq:p-atic_hydrodynamics}, \eqref{eq:psi}, \eqref{eq:p} and \eqref{eq:psi_1} represent the most generic hydrodynamic equations for $p-$atic liquid crystals with arbitrary discrete rotational symmetry. Yet, in various practical situations, the phase $\theta$ of the complex order parameter is the only hydrodynamic variable resulting from the broken rotational symmetry, whereas the scalar order parameter $\OP$ relaxes to its equilibrium value in a finite time. In the case of incompressible flows (i.e. $\nabla\cdot\bm{v}=0$), this occurs when $\theta$ varies over length scales much larger than
\begin{equation}\label{eq:mean_field_correlation_length}
\xi_{\rm m} = \sqrt{\frac{L}{|a_{2}|}}\;.
\end{equation}
Thus, in particular, in the absence of topological defects or other singular features, an example of which will be given in Sec. \ref{sec:flow_alignment} in the context of the so called flow tumbling instability. In compressible flows, this condition is further augmented by the requirement for the velocity field to be time-independent, as long as $\bar{\lambda}_{p}\ne 0$.

Under these circumstances, $\OP$ is uniform throughout the system and one can express the hydrodynamic equations in terms of the sole angle $\theta$:
\begin{equation}\label{eq:theta_p}
\frac{D\theta}{Dt} 
= \D\nabla^{2}\theta+\omega_{xy}
-|\Hp|\sin\left(p\theta-\Arg\Hp\right)\,,
\end{equation}
where $\D=K/\gamma$ is the rotational diffusion coefficient and the complex function $\Hp$, hereafter referred to as {\em flow alignment field}, embodies all the contributions arising from the interaction between $p-$atic order and flow. For $p \ge 2$, this can be expressed as
\begin{equation}\label{eq:flow_alignment_field}
\Hp = \frac{2}{p\OPeq}\left(\lambda_{p}\partial^{p-2}\U+\nu_{p}\partial^{p\,{\rm mod}\,2}\U^{\lfloor p/2 \rfloor}\right)\;.
\end{equation}
Similarly, for $p=1$, Eqs. \eqref{eq:psi_1} reduces to
\begin{equation}\label{eq:theta_1}
\frac{D\theta}{Dt} 
= \D\nabla^{2}\theta+\omega_{xy} 
- |\mathfrak{H}_{1}|\sin\left(2\theta-\Arg\mathfrak{H}_{1}\right)\,,
\end{equation}
with the flow alignment field given by:
\begin{equation}
\mathfrak{H}_{1} = \frac{\lambda_{1}}{\OPeq}\,\U\;.	
\end{equation}
It is worth noticing that $\mathfrak{H}_{1}$ and $\mathfrak{H}_{2}$ are formally identical. Thus, for $p=1$ and $2$, the angle $\theta$ obeys to the same hydrodynamic equation.

\subsection{\label{sec:stresses}Stresses in $p-$atics}

In order to complete the derivation of Eq. (\ref{eq:generic_hydrodynamics}b), one needs to calculate the reactive and viscous components of the stress tensor Eq. \eqref{eq:stress_tensor}. The reactive stress can be expressed as:
\begin{equation}\label{eq:reactive_stress}
\bm{\sigma}^{({\rm r})} = -P\mathbb{1}+\bm{\sigma}^{({\rm e})}+\bm{\sigma}^{({\rm d})}\;,	
\end{equation}
where $P$ is the pressure, $\bm{\sigma}^{({\rm e})}$ is the {\em elastic} stress, arising in response to static deformations of a fluid patch, and $\bm{\sigma}^{({\rm d})}$ is the {\em dynamic} stress originating from the reversible coupling between $p-$atic order and flow.

The elastic stress $\bm{\sigma}^{({\rm e})}$ can be calculated using the principle of virtual work (see e.g. Ref. \cite{Doi:1986}). This consists of equating the work performed by an arbitrary small deformation acting upon a generic fluid patch to the corresponding free energy variation. This procedure, reviewed in detail in Appendix \ref{app:stress}, yields:
\begin{equation}\label{eq:elastic_stress}
\sigma_{ij}^{({\rm e})} = - L\partial_{i}Q_{k_{1}k_{2}\cdots\,k_{p}}\partial_{j}Q_{k_{1}k_{2}\cdots\,k_{p}}\;,
\end{equation}
up to diagonal terms that can be incorporated into the pressure $P$.  

The dynamic contribution to the reactive stress, on the other hand, can be further decomposed into a symmetric part, arising from the {\em linear} flow alignment tensor $\bm{L}_{p}$, and an antisymmetric part, resulting from the corotational derivative in Eq. \eqref{eq:p-atic_hydrodynamics}. Both contributions can be calculated starting from the total entropy production rate (see e.g. Ref. \cite{Landau:1986}), which is given by
\begin{equation}\label{eq:entropy_production_1}
\dot{S} = \int \frac{{\rm d}^{2}r}{T}\,\left(\bm{\sigma}^{({\rm v})}:\nabla\bm{v}+\bm{H}_{p}\odot\frac{D\bm{Q}_{p}}{Dt}\right)\;.
\end{equation}
Taking $\bm{\sigma}^{({\rm v})}=\bm{\sigma}+P\mathbb{1}-\bm{\sigma}^{({\rm e})}$ and casting Eq. \eqref{eq:entropy_production_1} in the form of Eq. (\ref{eq:generic_hydrodynamics}d), yields:
\begin{align}\label{eq:dynamic_stress}
\sigma_{ij}^{({\rm d})} 
&=- \bar{\lambda}_{p} \bm{Q}_{p}\odot\bm{H}_{p}\,\delta_{ij} \phantom{\frac{p}{2}} \notag \\
&+ (-1)^{p-1}\lambda_{p}\partial^{p-2}_{k_{1}k_{2}\cdots\,k_{p-2}}H_{k_{1}k_{2}\cdots\,ij} \notag \\
&+ \frac{p}{2} \left( Q_{k_{1}k_{2} \cdots\,i} H_{k_{1}k_{2}\cdots\,j}-H_{k_{1}k_{2}\cdots\,i}Q_{k_{1}k_{2}\cdots\,j} \right)\;. 
\end{align}
More details about this calculation are given in Appendix \ref{app:stress}. As in nematic hydrodynamics, the second term on the right-hand side of Eq.~\eqref{eq:dynamic_stress}, originating from the correlational derivative of the tensor order parameter, is anti-symmetric by construction and, therefore, cannot equate the ensemble average of a microscopic stress tensor, which is symmetric. This symmetry property is, however, unimportant as the stress tensor enters in the momentum equation, Eq. (\ref{eq:generic_hydrodynamics}b), only via its divergence and it is always possible to construct a symmetric stress tensor, i.e. $\bm{\sigma}'$, such that $\nabla\cdot\bm{\sigma}'=\nabla\cdot\bm{\sigma}$. This procedure is reviewed, e.g., in Ref.~\cite{Landau:1986} for the case of nematics.

The non-linear flow alignment tensor $\bm{N}_{p}$, on the other hand, does not yield relevant contributions to the reactive stress. For even $\lfloor p/2 \rfloor$ values, i.e., for $p=4,\,5,\,8,\,9\ldots$, $\bm{N}_{p}$ is even under time reversal. It therefore describes an irreversible exchange of momentum between orientational degrees of freedom and flow. For odd $\lfloor p/2 \rfloor$ values, i.e., for $p=6,\,7,\,10,\,11\ldots$, the coupling is reversible, but, as nonlinear effects becomes relevant only when the shear rate is comparable to the inverse relaxation time (i.e. $\dot{\epsilon}\tau \approx 1$), their contribution to the total stress is negligible compared to the viscous stresses within the hydrodynamic regime. 

The viscous stress tensor $\bm{\sigma}^{({\rm v})}$, finally, can be expressed in the form:
\begin{equation}
\bm{\sigma}^{({\rm v})} = \bm{\eta}:\nabla\bm{v}\;,	
\end{equation}
by virtue of Onsager's reciprocal relations \cite{DeGroot:1984}. Here $\bm{\eta}$ is the rank$-4$ viscosity tensor, which is symmetric with respect to the first and second pair of indices, i.e. $i_{1} \leftrightarrow i_{2}$ and $i_{3} \leftrightarrow i_{4}$. In the absence of parity symmetry-breaking effects, such as odd viscosity \cite{Avron:1998} (which does not occur in passive liquid crystals, but could in driven or active chiral fluids, e.g. Ref.~\cite{Soni:2019}), it is also symmetric with respect to the exchange $\{i_{1}i_{2}\}\leftrightarrow\{i_{3}i_{4}\}$.  

Now, in the case of isotropic liquids, the viscosity tensor takes the standard form (see e.g. Ref. \cite{DeGroot:1984}):
\begin{equation}\label{eq:viscosity_tensor}
\bm{\eta}^{({\rm i})} = \zeta\mathbb{1}^{\otimes 2}+\eta \bm{\Delta}_{2,2}\;, 	
\end{equation}
with $\zeta>0$ and $\eta>0$ the bulk and shear viscosity respectively. By contrast, in $p-$atics, the viscosity tensor is augmented by an anisotropic component: i.e. $\bm{\eta}=\bm{\eta}^{(\rm i)}+\bm{\eta}^{({\rm a})}$, with $\bm{\eta}^{({\rm a})}$ a $p-$fold symmetric function of the director $\bm{n}$. Using standard  algebraic manipulations, it is possible to show that, with exception for $p=1,\,2$ and $4$, no combination of the anisotropic tensor $\traceless{\bm{n}^{\otimes p}}$ and the isotropic tensors $\bm{\mathbb{1}}$ and $\bm{\Delta}_{p,p}$ yields an anisotropic tensor that complies with the symmetry requirements of $\bm{\eta}$. Therefore, in these cases:
\begin{equation}\label{eq:viscous_stress}
\sigma_{ij}^{({\rm v})} = \zeta\tr(\bm{u})\,\delta_{ij} + 2 \eta \traceless{u_{ij}}\;. 
\end{equation} 
To illustrate this concept, let us consider, for instance, the case $p=3$. An additional contribution to the viscosity tensor could be obtained upon contracting $\traceless{\bm{n}^{\otimes 3}}$ with itself, i.e.: $\varrho_{3}\traceless{\bm{n}^{\otimes 3}}\cdot\traceless{\bm{n}^{\otimes 3}}$, 	
with $\varrho_{3}$ a constant. By virtue of Eq. \eqref{eq:q_dot_q}, however, this term is proportional to the isotropic tensor $\bm{\Delta}_{2,2}$, thus it affects the viscosity tensor by merely renormalizing the magnitude of the shear viscosity: $\eta\rightarrow\eta+\varrho_{3}/4$. Analogous arguments apply to other $p$ values.

For $p=1$ and $2$, however, it is possible to construct an anisotropic viscosity tensor $\bm{\eta}^{({\rm a})}$. In two dimensions, this consists of three independent viscosity coefficients, which, together with $\zeta$ and $\eta$, make a set of five independent viscosities (see e.g. Ref. \cite{Napoli:2016} for a general treatment that includes spatial curvature). Analogously, for $p=4$, one has:
\begin{equation}
\bm{\eta}^{({\rm a})} = \varrho_{4}\traceless{\bm{n}^{\otimes 4}}\;,	
\end{equation}
with $\varrho_{4} \sim \OPeq$ a constant, whose magnitude is constraint by the requirement $\dot{S}>0$, as demanded by the second law of thermodynamics. To make this constraint explicit, we calculate
\begin{equation}\label{eq:entropy_production_2}
\dot{S} 
= \hspace{-0.5ex} \int \frac{{\rm d}^{2}r}{T}
\left\{
  \eta\left|\traceless{\nabla\bm{v}}\right|^{2}
+ \zeta\left[\tr(\bm{u})\right]^{2}
+ \varrho_{4}\nabla\bm{v}\hspace{-0.25ex}:\hspace{-0.25ex}\traceless{\bm{n}^{\otimes 4}}\hspace{-0.25ex}:\hspace{-0.25ex}\nabla\bm{v}
\right\}\hspace{-0.25ex}.
\end{equation}
Then, switching again to the complex strain rate $\U$, defined in Eq. \eqref{eq:complex_strain_rate}, and taking advantage of the fact that $|\traceless{\nabla\bm{v}}|^{2}=2|\U|^{2}$ and
\begin{equation}
\nabla\bm{v}:\traceless{\bm{n}^{\otimes 4}}:\nabla\bm{v} = |\U|^{2}\cos\left(4\theta-\Arg\U\right)\;,
\end{equation}
allows one to express the tetratic entropy production as
\begin{equation}\label{eq:entropy_production_3}
\dot{S}
= \hspace{-0.5ex} \int \frac{{\rm d}^{2}r}{T}\, 
\left\{
\zeta[\tr(\bm{u})]^{2}
+ \left[2\eta+\varrho_{4}\cos(4\theta-2\Arg\U)\right]|\U|^2
\right\}\hspace{-0.25ex}.
\end{equation}
Finally, since either one of the two terms on the right-hand side of this equation can vanish independently and $-1\le\cos(4\theta-\Arg\U)\le 1$, $\dot{S}>0$ requires
\begin{equation}
-2\eta \le \varrho_{4} \le 2\eta\;.
\end{equation}
In summary, $p-$atic liquid crystals are expected to exhibit isotropic viscous stresses, except for polars (i.e. $p=1$), nematics (i.e. $p=2$) and tetratics (i.e. $p=4$), for which the orientational anisotropy affects viscous dissipation. Even in these three cases, however, the dissipational anisotropy is expected to become weaker at large length scales, owing to the fact that the viscosity coefficients appearing in $\bm{\eta}^{(\rm a)}$, which in turn are proportional to the order parameter (at least in mean field theory), are renormalized by thermal fluctuations and, therefore, vanish in the infinite system size limit. 

Although a full RG analysis (which we have not attempted here) is required in order to accurately assess the behavior of $\bm{\eta}^{(\rm a)}$ across different length scales, there are at least two reasons to expect the viscous anisotropy to be experimentally relevant. First, since the scalar order parameter, hence the anisotropic viscosities, decays as a power law in the presence of quasi-long-ranged order [see Eq. \eqref{eq:order_parameter_scale}], even macroscopically large samples could still exhibit appreciable anisotropy. For instance, assuming $\varrho_{4}/\eta \sim\OPeq\sim (a/\ell)^{\eta_{4}/2}$ (which is likely an overestimation, but the more accurate one can make without explicitly accounting for thermal fluctuations), taking $\eta_{4}=1/4$ and assuming the ultraviolet cut-off to be a molecular length scale, i.e., $a\approx 1$ nm, yields $\varrho_{4}/\eta\approx 0.13$ at a length scale $\ell=1$ cm. Thus even a centimeter-sized sample would exhibit an appreciable $13\%$ viscous anisotropy. This percentage is significantly larger for colloidal tetratics, such as those shown in Fig. \ref{fig:1}b, where $a \approx 1\,\mu$m and $\varrho_{4}/\eta \approx 0.32$ for $\ell=1$ cm. Second, as we will detail in Sec. \ref{sec:long_range_order}, subjecting the system to a finite shear rate induces long ranged order, which would make the anisotropy of the viscous tensor unambiguously measurable.

\section{\label{sec:backflow}Backflow effects}

As in other liquid crystals, the dynamics of the velocity field in $p-$atics is characterized by two different time scales, associated with  propagation of linear and angular momentum, i.e.
\begin{equation}
\label{eq:time_scales}
\tau_{\rm p} = \frac{\rho \ell^{2}}{\eta}\;,\qquad\tau_{\rm a} = \frac{\eta \ell^{2}}{K}\;.
\end{equation}
In turn, multiplying these by the shear rate $\deps$ yields two fundamental dimensionless numbers: the classic Reynolds number $\re=\deps\tau_{\rm p}$, proportional to the ratio of inertial to viscous forces, and the Ericksen number $\er=\deps\tau_{\rm a}$, proportional to the ratio of viscous to elastic torques (see e.g. Ref.~\cite{Kleman:2003}). As such, the latter quantifies the preponderance of an externally induced flow with respect to the internal backflow, namely the flow caused by spatial variations of $p-$atic order. Specifically, for $\er \gg 1$ ($\er \ll 1$), backflow effects are negligible (dominant). In a nematic film with thickness $w$, $\eta/w \approx 10\;{\rm mPa}\,{\rm s}$ and $K/w \approx 10\;{\rm pN}$ \cite{Kleman:2003}, taking $\ell \approx 1\,{\rm mm}$ and $\deps \approx 10\,{\rm s}^{-1}$ gives $\er \approx 10^{4}$. Thus, at the macroscopic scale, it is generally possible to neglect backflow, except in proximity to boundary layers or topological defects, where the local orientation can vary over submicron distances. At the microscopic scale, on the other hand, backflow effects are more prominent and thermal fluctuations can temporarily disrupt the condition $\er\gg 1$, even if this is fulfilled at the scale of the system size. 

In this Section, we demonstrate that, in the Stokesian limit, that is when inertial effects are negligible, and for $\er\approx 1$, backflow effectively enhances rotational diffusion and can be accounted for by replacing
\begin{equation}\label{eq:effective_diffusion_coefficient}
\D \to \De = K\left(\frac{1}{\gamma}+\frac{1}{4\eta}\right)\;,	
\end{equation}
in Eqs. \eqref{eq:theta_p} and \eqref{eq:theta_1}. To prove this statement we observe that, in the Stokesian limit, Eq. (\ref{eq:generic_hydrodynamics}b) reduces to
\begin{subequations}\label{eq:stokes}
\begin{gather}
\eta\nabla^{2}\bm{v}-\nabla P + \nabla\cdot\bm{\sigma}^{({\rm r})} = \bm{0}\;,\\[5pt]
\nabla\cdot\bm{v} = 0\;,
\end{gather}
\end{subequations}
where $\bm{\sigma}^{({\rm r})}$ is reactive stress tensors defined in Sec. \ref{sec:stresses}. Under the assumption of homogeneous scalar order parameter, this can be cast in the classic form given in Ref. \cite{Zippelius:1980a}, namely
\begin{equation}\label{eq:theta_stress} 
\bm{\sigma}^{({\rm r})} = -P\mathbb{1}+\frac{K}{2}\,\bm{\varepsilon}\nabla^{2}\theta-K \nabla\theta \otimes \nabla\theta\;, 
\end{equation}
where $\bm{\varepsilon}$ is again the antisymmetric tensor defined in Sec. \ref{sec:order_parameter}. Now, a simple solution of Eqs. \eqref{eq:stokes} can be obtained by decomposing the velocity field in an externally driven component, $\bm{v}^{({\rm e})}$, and a backflow component, $\bm{v}^{({\rm b})}$, so that
\begin{equation}
\bm{v} = \bm{v}^{({\rm e})}+\bm{v}^{({\rm b})}\;.	
\end{equation}	
For simplicity, here we take $\bm{v}^{(\rm e)}=\bm{0}$ and assume the flow is solely due to backflow effects. This hypothesis will be lifted in the following Section. Then, substituting Eq.~\eqref{eq:theta_stress} in Eq.~\eqref{eq:stokes} and approximating all the fields at the linear order in $\nabla\theta$, readily yields
\begin{subequations}\label{eq:backflow}
\begin{gather}
\bm{v}^{({\rm b})} = - \frac{K}{2\eta}\,\bm{\varepsilon}\cdot\nabla\theta+\mathcal{O}\left(|\nabla\theta|^{2}\right)\;,\\[5pt]
P = P_{0}+\mathcal{O}\left(|\nabla\theta|^{2}\right)\;,
\end{gather}
\end{subequations}
with $P_{0}$ a uniform pressure. Thus, away from the boundary, spatial variations in the average orientation $\theta$ drive a transverse backflow, whose strain rate and vorticity can be approximated from Eqs. \eqref{eq:backflow} as
\begin{subequations}\label{eq:backflow_u_omega}
\begin{gather}
u_{xx} = - u_{yy} \approx -\frac{K}{2\eta}\,\partial_{xy}^{2}\theta\;,\\
u_{xy} = u_{yx} \approx \frac{K}{4\eta}\,(\partial_{x}^{2}-\partial_{y}^{2})\theta\;,\\
\omega_{xy} = -\omega_{yx} \approx \frac{K}{4\eta}\,\nabla^{2}\theta\;.	
\end{gather}
\end{subequations}
Finally, using Eqs. \eqref{eq:backflow_u_omega} in Eq. \eqref{eq:theta_p} and truncating the latter equation at the linear order in $\nabla\theta$ gives
\begin{equation}\label{eq:theta_backflow}
\partial_{t}\theta = \De\nabla^{2}\theta-|\Hp|\sin(p\theta-\Arg\Hp)\;,	
\end{equation}
with $\De$ the effective rotational diffusion coefficient defined in Eq. \eqref{eq:effective_diffusion_coefficient}. Analogously, the flow alignment field is given by
\begin{equation}
\Hp = \frac{i\lambda_{p}}{\eta}\frac{K}{p\OPeq}\,\partial^{p}\theta\;.	
\end{equation} 
Thus, in the absence of an externally driven flow and strong distortion of the local orientation, backflow has the effect of speeding up the relaxational dynamics of the $p-$atic director by increasing the effective rotational diffusion coefficient, but ultimately leads to a homogeneous and stationary configuration, where $\theta={\rm const}$ and $\bm{v}^{({\rm b})}=\bm{0}$, unless the boundary conditions demand otherwise.

To conclude, we stress that the above derivation is rooted in three important simplifying assumptions. First, inertial effects are negligible and the velocity field can be found within the Stokesian limit. Second, viscous and elastic stresses are comparable in magnitude. Third, the $p-$atic director gently varies across the system. In terms of the previously defined Reynolds ($\re$) and Ericksen ($\er$) numbers, the first two assumptions imply $\re \ll 1$ and $\er \approx 1$, or, equivalently
\begin{equation}\label{eq:re_vs_er}
\frac{\re}{\er} = \frac{\rho K}{\eta^{2}} \ll 1\;.	
\end{equation}
In most of thermotropic liquid crystals, $\re/\er \approx 10^{-4}$ at room temperature and Eq. \eqref{eq:re_vs_er} is well satisfied \cite{Kleman:2003,Mazenko:1983}. Furthermore, in colloidal $p-$atics (see Fig. \ref{fig:1}), as one cools the sample down towards the liquid-solid phase transition, both the shear viscosity $\eta$ and the orientational stiffness $K$ are predicted to diverge like $\xi_{p}^{2}$ \cite{Zippelius:1980a}, where $\xi_{p} \sim \exp(bt^{-\nu_{p}})$ is the correlation length, with $b$ a constant of order one, $t=(T-T_{\rm m})/T_{\rm m}$, with $T_{\rm m}$ the melting temperature, and $\nu_{p} = 1/2$ for all $p$ values \cite{Nelson:1978,Ostlund:1981,Radzihovsky:unpublished} expect $p=6$, for which $\nu_{6} \approx 0.36963$ \cite{Halperin:1978,Nelson:1979,Young:1979}. Thus $\re/\er \to 0$ as the liquid-solid phase transition is approached from above. The third assumption, on the other hand, requires $|\nabla\theta| \approx d^{-1}$, with $d$ the system size. Since $|\bm{v}| = K/(2\eta)|\nabla\theta|$ and, away from topological defects, $\er = \eta v d/K$, this assumption translates once again into the requirement $\er \approx 1$, thus it is already accounted for in Eq. \eqref{eq:re_vs_er}.

\section{\label{sec:long_range_order}Long-range order in $p-$atics under shear}

As we discussed in Sec. \ref{sec:order_parameter}, two-dimensional $p-$atics do not, in fact, exhibit long-ranged orientational order in equilibrium. Rather, orientational order is quasi-long-ranged \cite{Halperin:1978,Nelson:1979}, that is, the orientational correlation function, Eq. \eqref{eq:correlation_function}, decays to zero as a power law as the spatial separation $|\bm{r}| \rightarrow \infty$, with a non-universal exponent, as shown explicitly in Eqs. \eqref{eq:correlation_function} and \eqref{eq:eta}. This implies a vanishing order parameter as well. The latter can be calculated by taking the long distance limit of the correlation function:
\begin{equation}\label{eq:order_parameter_limit}
\lim_{|\bm{r}| \rightarrow \infty} \left\langle \psi_{p}^{*}(\bm{r})\psi_{p}(\bm{0}) \right\rangle
= \left\langle \psi_{p}^{*}(\bm{r}) \right\rangle \left\langle \psi_{p}(\bm{0}) \right\rangle 
= \OP^{2}\;.
\end{equation}
Thus, in the thermodynamic limit, the $p-$atic order parameter vanishes as demanded by Eq. \eqref{eq:order_parameter_scale}. At equilibrium, this classic result can be recovered starting from the ${\rm O}(2)$ Hamiltonian 
\begin{equation}\label{eq:xy_hamiltonian}
\mathcal{H}=\frac{1}{2}\,K\int {\rm d}^{2}r\,|\nabla\vartheta|^{2} \,,
\end{equation}
from which one can calculate
\begin{equation}\label{eq:psipsi}
\langle \psi_{p}^{*}(\bm{r})\psi_{p}(\bm{0}) \rangle = e^{-p^{2}g(\bm{r})}\;,
\end{equation}
where $g(\bm{r})$ is the connected correlation function of the microscopic orientation $\vartheta$: 
\begin{equation}\label{eq:connected_correlation_function}
g(\bm{r}) 
= \frac{1}{2}\left\langle[\vartheta(\bm{r})-\vartheta(\bm{0})]^{2}\right\rangle\;.
\end{equation}
Applying the equipartition theorem to Eq. \eqref{eq:xy_hamiltonian}, one can readily show that
\begin{equation}\label{eq:theta_variance}
g(\bm{r})
= \frac{k_{B}T}{K} \int_{0<|\bm{q}|<\Lambda}\frac{{\rm d}^{2}q}{(2\pi)^{2}}\,\frac{1-e^{i\bm{q}\cdot\bm{r}}}{q^{2}}\;, 
\end{equation}
where $\Lambda = 2\pi/a$. This leads to the asymptotic result
\begin{equation}\label{eq:g_no_flow}
g(\bm{r}) \approx \frac{k_{B}T}{2\pi K} \log \frac{|\bm{r}|}{a}\;, \qquad |\bm{r}|\gg a\;,
\end{equation}
from  which one readily obtains Eq. \eqref{eq:correlation_function}, with the exponent $\eta_{p}$ given by Eq. \eqref{eq:eta} (see e.g. Ref. \cite{Chaikin:1995}). 

In this Section, we show that an externally imposed uniform shear induces long-ranged order. Intuitively, this can be understood by observing that hydrodynamic flow introduces a time scale $\tau_{\rm s}=1/\deps$, with $\deps$ the typical shear rate of the flow, as well as the length scale
\begin{equation}\label{eq:ls}
\ls = \sqrt{\frac{\De}{\deps}}\;.
\end{equation}
The latter, hereafter referred to as the {\em shear length scale}, is the distance at which elastic and hydrodynamic torques balance each other. As a consequence, fluctuations are highly anisotropic, but are suppressed at length scales larger than $\ls$, with respect to their equilibrium counterpart. Thus, it is the shear length $\ls$, rather than the system size, that provides the long wavelength (i.e. infrared) cutoff on the Goldstone modes and one can expect
 \begin{equation}\label{eq:order_parameter_epsilon}
\OP \sim \left(\frac{a}{\ell_{\rm s}}\right)^{\eta_{p}/2} \sim (\deps\tau)^{\,\eta_{p}/4}\;,
\end{equation}
where $\tau = a^2/\De$ is the characteristic relaxation time at the cut-off length scale. Furthermore, since $\eta_{p}<1/4$, Eq. \eqref{eq:order_parameter_epsilon} implies that even a very small shear rate can induce large, i.e. $\mathcal{O}(1)$, order parameter values. This upper bound also entails important physical consequences for flow alignment, as we will see in Sec. \ref{sec:flow_alignment}. 

The mechanism illustrated above is analogous to that described by Onuki and Kawasaki in the context of generic second order phase transitions \cite{Onuki:1979a} and latter invoked to account for the solid-like behavior of smectic layers \cite{Ramaswamy:1984}. In the following, we will demonstrate through a detailed calculation that the same mechanism results in the suppression of Goldstone modes in two-dimensional liquid crystals under shear, thereby promoting quasi-long-ranged into long-ranged order.

\subsection{\label{sec:linear_theory} Linear theory}  

\begin{figure}[t]
\centering
\includegraphics[width=\columnwidth]{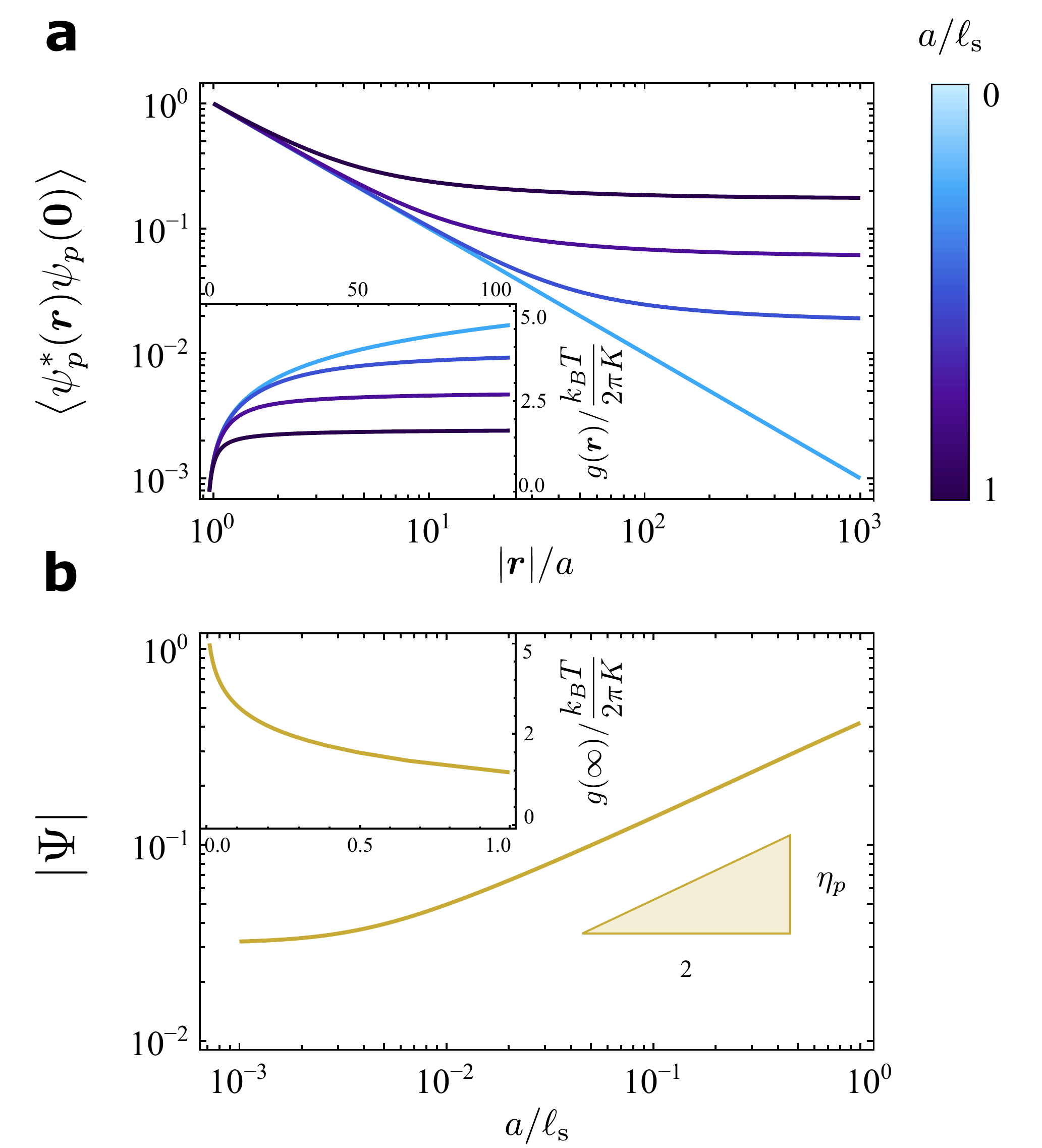}	
\caption{\label{fig:correlation_function} (a) Two-point $p-$atic correlation function, as defined in Eq. \eqref{eq:correlation_function}, versus distance for various shear rates expressed in terms of the dimensionless ratio $a/\ell_{\rm s}$, with $a$ a short distance cut-off and $\ell_{\rm s}$ the shear length scale defined in Eq. \eqref{eq:ls} for $\phi=0$. Inset: the connected correlation function $g=g(\bm{r})$, Eq. \eqref{eq:g}, versus distance. (b) $p-$atic order parameter $\OP$ versus shear rate, expressed in terms of $a/\ell_{\rm s}$. Inset: the asymptotic value $g(\infty) = \lim_{|\bm{r}|\rightarrow\infty}g(\bm{r})$.}
\end{figure}

In this Subsection we consider an incompressible $p-$atic liquid crystal subject to thermal fluctuations and to an externally imposed shear flow. This can be achieved by
augmenting hydrodynamic equations for $p-$atic phase $\vartheta$ and vorticity $\omega=2\omega_{xy}=\partial_{x}v_{y}-\partial_{y}v_{x}$ with additional random fields. At the linear order in $\vartheta$, this gives
\begin{subequations}\label{eq:fluctuating_hydrodynamics}
\begin{gather}
\rho (\partial_{t}+\bm{v}\cdot\nabla)\omega = \eta \nabla^{2}\omega+\nabla_{\perp}\times\nabla\cdot\bm{\sigma}^{({\rm r})}+\xi^{({\rm \omega})}\;,\\[5pt]
(\partial_{t}+\bm{v}\cdot\nabla)\vartheta = \D\nabla^{2}\theta+\frac{\omega}{2}+\xi^{({\rm \vartheta})}\;.
\end{gather}
\end{subequations}
with  $\nabla_{\perp}=\bm{e}_{z}\cdot\nabla$. The random fields $\xi^{(\omega)}=\xi^{(\omega)}(\bm{r},t)$ and $\xi^{(\vartheta)}=\xi^{(\vartheta)}(\bm{r},t)$ have zero mean, are Gaussianly distributed and their correlation functions are consistent with the fluctuation-dissipation theorem, so that
\begin{subequations}\label{eq:generic_noise_correlation}
\begin{multline}
\left \langle \xi^{(\alpha)}(\bm{r},t)\xi^{(\beta)}(\bm{r}',t') \right\rangle\\	
= 2k_{\rm B}T\left(\frac{1}{\gamma}\,\delta_{\alpha\vartheta}\delta_{\beta\vartheta}+\eta\delta_{\alpha\omega}\delta_{\beta\omega}\nabla^{4}\right)\delta(\bm{r}-\bm{r}')\delta(t-t')\;.
\end{multline}
\end{subequations}
To make progress, we decompose the velocity field into an average and a fluctuating component:
\begin{equation}\label{eq:velocity_decomposition}
\bm{v}= \langle\bm{v}\rangle+\delta\bm{v}\;,
\end{equation}
so that $\langle \delta\bm{v} \rangle = \bm{0}$. To compute the average velocity, we consider a simple shear flow generated by placing the sample between parallel plates and sliding them over each other at constant relative velocity. Taking the plates parallel to the $x-$direction, yields
\begin{equation}\label{eq:simple_shear}
\langle \bm{v} \rangle = \dot{\epsilon}y\bm{e}_{x}\;,
\end{equation}
with $\deps$ a constant shear-rate. Furthermore, we assume the system in the regime discussed in Sec. \ref{sec:backflow} and subject to the constraint expressed by Eq. \eqref{eq:re_vs_er}, so that the backflow effects can be incorporated directly into the rotational diffusion coefficient, Eq. \eqref{eq:effective_diffusion_coefficient}. Analogously, as we detail in Appendix \ref{app:long_ranged_order}, the random field $\xi^{(\omega)}$ results in a renormalization of the orientational noise $\xi^{(\vartheta)}$, so that Eqs. \eqref{eq:fluctuating_hydrodynamics} can be reduced, at the linear order in all the fluctuating fields, to a single stochastic partial different equation:
\begin{equation}\label{eq:theta_simplified_1}
\partial_{t} \vartheta + \dot{\epsilon}y\,\partial_{x}\vartheta = \De\nabla^{2}\vartheta -\frac{\dot{\epsilon}}{2} + \xi\;,
\end{equation}
where we have used the fact that the vorticity arising from the externally imposed field Eq. \eqref{eq:simple_shear} is given by $\omega=-\dot{\epsilon}$. The effective rotational diffusion coefficient is given by Eq. \eqref{eq:effective_diffusion_coefficient}, whereas $\xi=\xi(\bm{r},t)$ is the effective orientational noise field, whose correlation function is given by
\begin{equation}\label{eq:noise_correlation}
\left\langle\xi(\bm{r},t)\xi(\bm{r}',t')\right\rangle = \frac{2k_{\rm B}T}{\Ge}\,\delta(\bm{r}-\bm{r}')\delta(t-t')\;,
\end{equation}
with $\Ge=K/\De$. In Appendix \ref{app:long_ranged_order} we formally solve Eq. \eqref{eq:theta_simplified_1} to express the orientational field $\vartheta(\bm{q}, t)$ in Fourier space as a linear functional of the spatially Fourier transformed noise $\xi(\bm{q},t)$. We can then autocorrelate this expression with itself and use Eq. \eqref{eq:noise_correlation} for the noise correlations to obtain an expression for the equal time correlation $\langle |\vartheta(\bm{q},t)|^{2} \rangle$. Then computing
\begin{equation}\label{eq:q_integral_1}
g(\bm{r}) = \lim_{t\rightarrow\infty}\int_{0<|\bm{q}|<\Lambda}\frac{{\rm d}^{2}q}{(2\pi)^{2}}\,(1-e^{i\bm{q}\cdot\bm{r}})\langle |\vartheta(\bm{q},t)|^{2} \rangle\;,
\end{equation}
where $\langle |\vartheta(\bm{q},t)|^{2} \rangle$ is an orientational structure factor defined from the relation
\begin{equation}\label{eq:t_integral}
\left\langle \vartheta(\bm{q},t)\vartheta(\bm{q}',t)\right\rangle = (2\pi)^{2} \langle |\vartheta(\bm{q},t)|^{2} \rangle \delta(\bm{q}+\bm{q}')\delta(t-t')\;,
\end{equation}
we obtain
\begin{equation}\label{eq:g}
g(\bm{r}) = \frac{k_{B}T}{2\pi K}\int_{0}^{\infty}{\rm d}\tau\,\frac{e^{-\mathcal{G}(\tau,\phi)\left(\frac{a}{\ls}\right)^{2}}-e^{-\mathcal{G}(\tau,\phi)\left(\frac{|\bm{r}|}{\ls}\right)^{2}}}{\tau\sqrt{4+\frac{1}{3}\tau^{3}}}\;,
\end{equation}
where we have defined 
\begin{equation}\label{eq:phi}
\mathcal{G}(\tau,\phi) = \frac{1-\frac{1}{2}\tau\sin 2\phi+\frac{1}{3}\tau^{2}\sin^{2}\phi}{2\tau\left(4+\frac{1}{3}\tau^{2}\right)}\;.
\end{equation} 
Fig. \ref{fig:correlation_function}a (inset) shows a plot of the connected correlation function versus $|\bm{r}|/a$ for various $a/\ls$ values. For $a/\ls\rightarrow 0$, corresponding to $\deps \to 0$, this displays the characteristic logarithmic growth of $p-$atics at equilibrium. By contrast, for $a/\ell_{\rm s}>0$ the connected correlation function does not grow without bound, but rather plateaus at a finite value. Recalling Eq. \eqref{eq:psipsi}, this implies that $\langle \psi_{p}^{*}(\bm{0})\psi_{p}(\bm{r})\rangle$ converges to a finite value at large scales (Fig. \ref{fig:correlation_function}a), indicating that a shear flow of arbitrary finite shear rate render the orientational order of $p-$atic phases long-ranged. The corresponding order parameter $\OP$, given by Eq. \eqref{eq:order_parameter_limit}, can be calculated form the asymptotic value of the $p-$atic correlation function (Fig. \ref{fig:correlation_function}b inset) and is plotted in Fig. \ref{fig:correlation_function}b versus $a/\ell_{\rm s}$.

To make this result more explicit, one can approximate the connected correlation function, Eq. \eqref{eq:g}, at short and long distances.
The result is:
\begin{equation}\label{eq:g_limit}
g(\bm{r})
\approx \frac{k_{B}T}{2\pi K}
\left\{
\begin{array}{lll}
\log \frac{|\bm{r}|}{a} & & |\bm{r}| \ll \ell_{\rm s} \; \\[10pt]
G_{0} - \frac{1}{2}\Ei\left(-\frac{a^{2}}{8\ell_{\rm s}^{2}}\right)\; & & |\bm{r}| \gg \ell_{\rm s}\;,
\end{array}
\right.	
\end{equation}	
where $\Ei$ is the exponential integral and $G_{0}=1/2\,\arcsinh 2\sqrt{3} \approx 0.9779$ (see Appendix \ref{app:long_ranged_order} for details). Thus, as already evident from the Fig. \ref{fig:correlation_function}a, the short distance behavior of the correlation function is unaffected by the shear flow, as a consequence of the fact that, well below the shear length scale $\ell_{\rm s}$, the fluctuations of the $p-$atic orientation $\vartheta$ are mainly governed by the competition between thermal and elastic torques. By contrast, at distances much larger than $\ell_{\rm s}$, elastic torques are outweighed by hydrodynamic torques, resulting in the emergence of global alignment. Using Eqs. \eqref{eq:order_parameter_limit} and \eqref{eq:g_limit} and the expansion of the exponential integral given in Appendix \ref{app:long_ranged_order}, we recover the expression for the order parameter given in Eq. \eqref{eq:order_parameter_epsilon}. The latter, in turn, vanishes for $\dot{\epsilon}\rightarrow 0$, when $\ell_{\rm s}\rightarrow\infty$, thereby recovering the equilibrium absence of long-ranged order. 

\begin{figure}[t]
\centering
\includegraphics[width=\columnwidth]{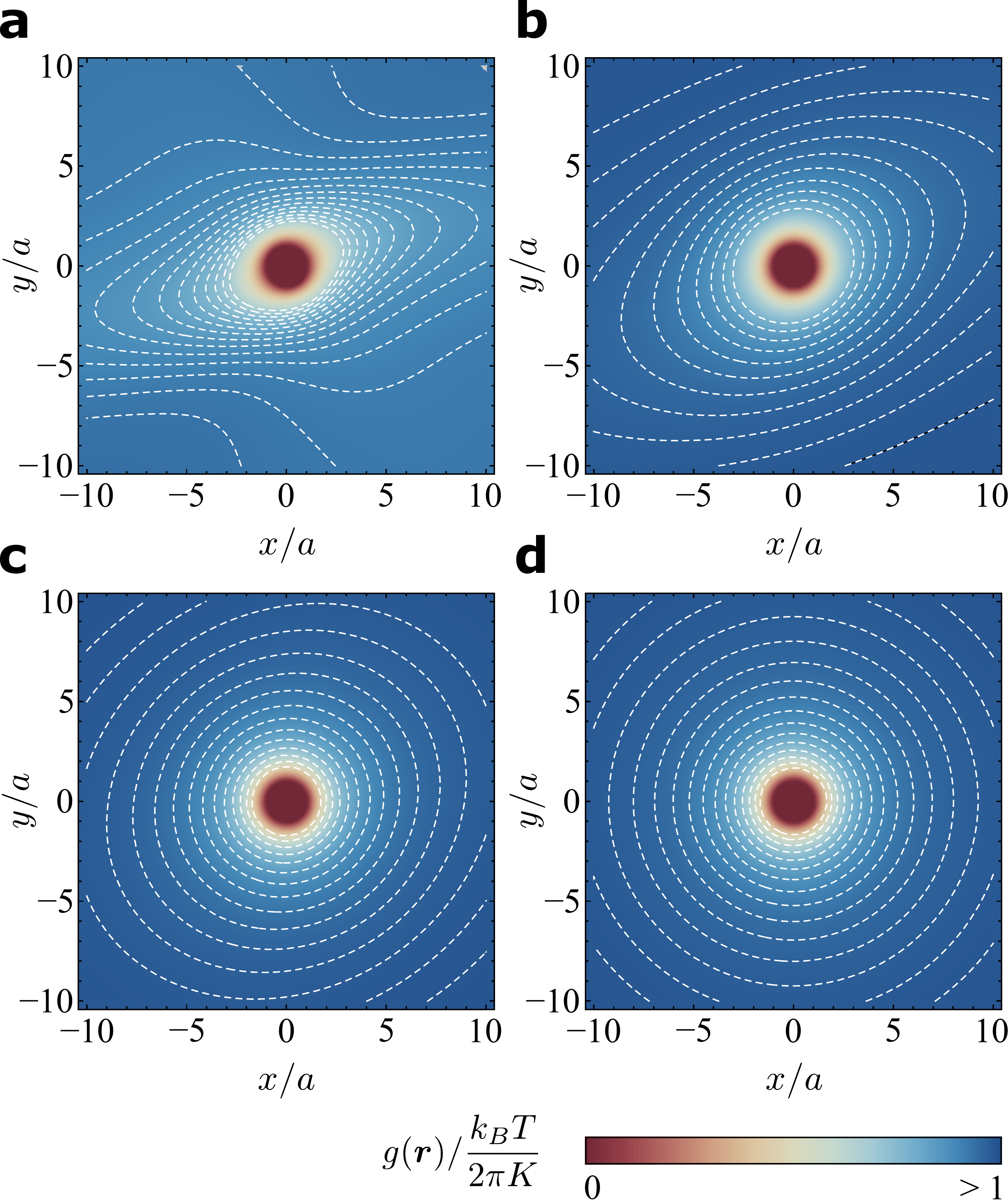}
\caption{\label{fig:contour_plot} Density plot of the connected correlation function, as defined in Eq. \eqref{eq:connected_correlation_function} as a  function of position for (a) $a/\ell_{\rm s}=1$, (b) $a/\ell_{\rm s}=1/3$, (c) $a/\ell_{\rm s}=1/10$ and (d) $a/\ell_{\rm s}=0$.}
\end{figure}

The inherent anisotropy of the shear flow, Eq. \eqref{eq:simple_shear}, has the further effect of rendering the orientational correlation of the $p-$atic anisotropic, as can be seen from the $\phi-$dependence in Eq. \eqref{eq:g} and the contour plots shown in Fig. \ref{fig:contour_plot}. Nevertheless, as it is clear from Eq. \eqref{eq:g_limit}, this effect disappears at both small and large scales.

In summary, to leading order in the externally imposed shear rate $\deps$, the effect of such shear is to induce long-ranged order, as manifest by a non-zero value of $\OP$ given by Eq. \eqref{eq:order_parameter_epsilon}. Although the demonstration presented here is strictly valid only in the subset of parameter space described by Eq. \eqref{eq:re_vs_er}, where backflow effects can be accounted for via a simple redefinition of the rotational diffusion coefficient, we expect this result to carry over to other regimes, provided the longest relaxation time in the dynamics of $\vartheta$, i.e. $\tau_{\max}$, is larger than the time scale of the externally applied shear flow: i.e. $\tau_{\max}>1/\deps$. In these circumstances, and analogously to the regime discussed here, the orientational fluctuations are expected to be suppressed by the flow at length scales larger than $\ell_{\max}=\sqrt{\tau_{\max}\D}$. Finally, Eq. \eqref{eq:theta_simplified_1}, implies that the phase $\theta$ of the complex order parameter $\Psi_{p}$, though coherent in space due to the suppression of fluctuations by the imposed shear, is not fixed in time, but rather rotates at a constant rate $-\deps/2$. Such a state is analogous to the ``tumbling'' state found in nematics for $\lambda_{2}<1$ \cite{Kleman:2003}. In Sec. \ref{sec:flow_alignment} we will further elaborate on this tumbling state and its onset in confined systems and we will show that, as a consequence of the nonlinear couplings between orientation and flow, Eq. \eqref{eq:n_tensor}, it is possible to obtain a flow aligned state, but only at sufficiently high shear rates $\deps$.

\subsection{\label{sec:rg} Nonlinear theory}

In this Subsection, we perform a simple RG analysis to assess the validity of the preceding linear theory, which ignored nonlinear flow-alignment effects and we will see that there is a surprising connection between this problem and the dynamics of the roughening transition \cite{Chiu:1978}.

For odd $p \ne 1$ values, $\Hp=0$, by virtue of the fact that $\deps$ is uniform throughout the system and the nonlinear terms cancel identically. For even $p$ values, on the other hand, including nonlinearities yields the following hydrodynamic equation for the fluctuating field $\vartheta$:
\begin{equation}\label{eq:nonlinear_eom}
\partial_{t}\vartheta + \dot{\epsilon}y\,\partial_{x}\vartheta 
= \De\nabla^{2}\vartheta-\frac{\dot{\epsilon}}{2} 
- h_{0} \sin p\left(\vartheta-\frac{\pi}{4}\right)+\xi \;.
\end{equation}
where $h_{0}$ is the ``bare'' amplitude of the flow alignment field $\Hp$ and is given by
\begin{equation}\label{eq:h_bare}
h_{0} = \left(\frac{\deps}{2}\right)^{p/2}
\left\{
\begin{array}{lll}
\frac{\lambda_{2}}{\OPeq} & & p=2 \\[10pt]
\frac{2\nu_{p}}{p\OPeq} & & p=4,\,6,\,8\ldots
\end{array}
\right.	
\end{equation}
Next, performing the transformation $\vartheta \to \vartheta+\pi/4$, and ignoring the terms resulting from convection and vorticity, i.e. $\dot{\epsilon}y\,\partial_{x}\vartheta$ and $\dot{\epsilon}/2$, which do not affect the dynamics of the local orientation at scales $\ell\ll\ls$, we can rewrite Eq. \eqref{eq:nonlinear_eom} as:
\begin{equation}\label{eq:Sine_Gordon}
\partial_{t}\vartheta = \De\nabla^{2}\vartheta - h_{0}\sin p\vartheta+\xi \;.
\end{equation} 
This equation is simple relaxational model for a sine-Gordon theory and, following Ref.~\cite{Chiu:1978}, can be analyzed using dynamical RG in order to obtain the following equations describing how the parameters $h=h(\ell)$, $\De=\De(\ell)$ and $K=K(\ell)$, change at the length scale $\ell>a$. This gives:
\begin{subequations}\label{eq:rg}
\begin{gather}
\frac{{\rm d}h}{{\rm d}l} = h\left[2-\frac{\eta_p}{2}+\mathcal{O}\left(h\tau\right)\right]\;,\\[5pt] 
\frac{{\rm d}\De}{{\rm d}l} = \mathcal{O}\left(h^{2}\tau^{2}\right)\;,\\[5pt]
\frac{{\rm d}K}{{\rm d}l} = \mathcal{O}\left(h^{2}\tau^{2}\right)\;,
\end{gather}
\end{subequations}
with $l = \log(\ell/a)$ and $\tau$ as in Eq. \eqref{eq:order_parameter_epsilon}. From Eqs. (\ref{eq:rg}b,c), we see that, as long as
\begin{equation}\label{eq:rg_inequality}
h \tau \ll 1\;,
\end{equation}
both $\De$ and $K$ are not renormalized by fluctuations and $\tau$ equates the time scale of the rotational dynamics at the length scale of the ultraviolet cut-off $a$. Thus, the right-hand side of Eq. (\ref{eq:rg}a) is constant and the equation can be immediately integrated to give
\begin{equation}\label{eq:rg_solution}
h(\ell)=h_{0}\left(\frac{\ell}{a}\right)^{2-\eta_p/2} \;. 
\end{equation}
Now, Eq. \eqref{eq:rg} holds for length scales $\ell < \ls$. For $\ell>\ls$, on the other hand, the terms resulting from convection and vorticity in Eq. \eqref{eq:nonlinear_eom} become important and, as shown earlier, cut off thermal fluctuations at the large scale. Thus, fluctuations no longer renormalize the material parameters at any length scale larger than $\ls$ and the linear theory is again valid, unless the coupling $h$ itself has by then become so large as to violate Eq. \eqref{eq:rg_inequality}. To exclude this possibility one can compute the renormalized coupling at the crossover scale. Using Eqs. \eqref{eq:ls} and \eqref{eq:rg_solution} gives
\begin{equation}
h\tau \sim \deps^{\,p/2-1+\eta_{p}/4}\;,
\end{equation}
which vanishes for small $\deps$ values, provided
\begin{equation}
p>2-\frac{\eta_{p}}{2} \,.
\label{cond2}
\end{equation}
Since $\eta_p>0$, this condition is obviously satisfied, meaning that the shear flow term is {\em irrelevant} at small shear rates, for all $p\ge2$. The same argument applies to the case $p=1$, which, as we explained in Sec. \ref{sec:theta}, is formally identical to $p=2$. Using again Eq. \eqref{eq:rg_solution}, the condition Eq. \eqref{eq:rg_inequality} requires then
\begin{equation}
h\tau \sim \deps^{\,\eta_{1}/4}\;,	
\end{equation}	
which again vanishes for small shear rates.

\section{\label{sec:flow_alignment}Flow alignment in channel and Taylor-Couette flows}

In this Section we demonstrate that the nonlinear couplings between $p-$atic order and flow, embodied by the field $\Hp$ in Eq. \eqref{eq:theta_p}, although they can not lead to flow alignment at small shear rates, could potentially do so at high shear rates. This possibility was missed by previous hydrodynamic theories of $p-$atics, because of the continuous, i.e. ${\rm O}(2)$ rotational symmetry. 

Specifically, we will discuss two classic examples of liquid crystals hydrodynamics: a generic $p-$atic liquid crystal confined in an infinitely long channel whose upper wall is dragged at constant speed (Sec. \ref{sec:simple_shear} and Fig. \ref{fig:5}a) as well as a two-dimensional analog of a Taylor-Couette cell, consisting of a annulus delimited by two counter-rotating walls (Sec. \ref{sec:taylor_couette} and Fig. \ref{fig:5}b). In both cases, we assume the $p-$atic fluid incompressible (i.e. $\nabla\cdot\bm{v}=0$) and strongly anchored to the lateral walls.

\subsection{\label{sec:simple_shear}Channel flow}

\begin{figure*}[t]
\centering
\includegraphics[width=\textwidth]{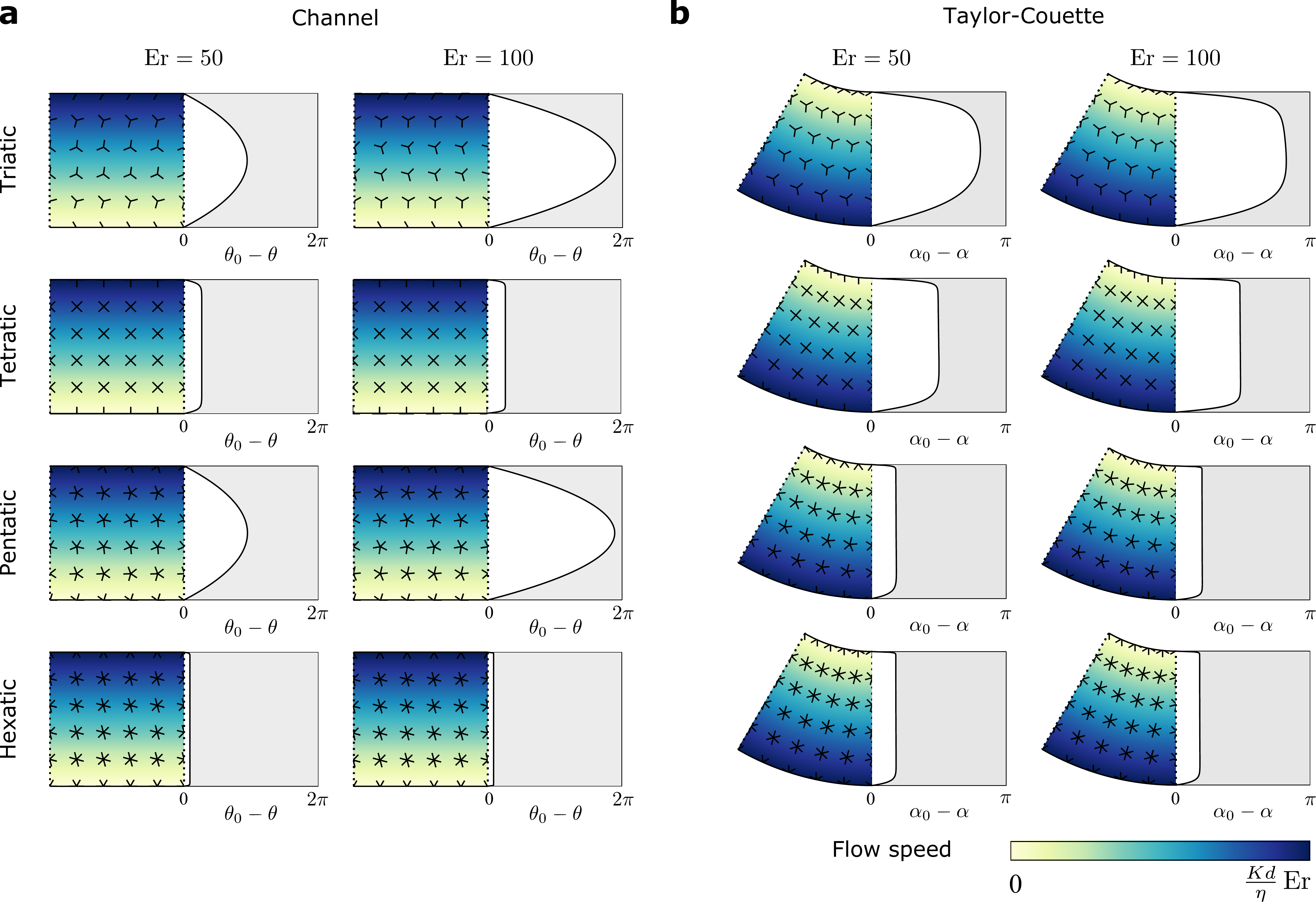}
\caption{\label{fig:5}Examples of high Ericksen number $p-$atic flow in a channel (a) and a Taylor-Couette cell (b), when $\xi_{\rm m} \rightarrow 0$ and the scalar order parameter can be assumed uniform throughout the system, i.e. $\OP=\OPeq$. (a) Numerical solution of Eq. \eqref{eq:leslie} for triatics ($p=3$), tetratics ($p=4$), pentatics ($p=5$) and hexatics ($p=6$) with boundary conditions $\theta_{0}=\Delta\theta=0$, with $d$ the channel thickness. The left-hand side of all plots shows the configuration of the $p-$atic director, represented by $p-$headed stars, superimposed to a heat map of the flow speed. The solid lines denote the channel walls, whereas the dotted lines mark the position of the channel inlet/outlet. The left-hand side of the plots shows the configuration of the $p-$atic director in terms of the angle $\theta_{0}-\theta$. (b) Numerical solutions of Eq. \eqref{eq:alpha_steady} with boundary conditions $\alpha_{0}=\pi/2$ and $\Delta\alpha=0$. In all plots the parameter values are $\tau_{\rm p}/\tau_{\rm a}=1$, $\lambda_{p}/d^{p-2}=1.5$ and $\nu_{p}/(d^{p\,{\rm mod}\,2}\tau_{\rm p}^{\lfloor p/2\rfloor-1})=2.0$. For the plots in panel (b) we set $R_{1}/d=1$, $R_{2}/d=2$ and $\Omega_{1}=0$.}
\end{figure*}

Let us consider a $p-$atic liquid crystal confined within a two-dimensional channel of infinite length along the $x-$direction and finite width $d$. The upper wall is dragged at speed $v_{0}$, in such a way that $\er=\eta v_{0} d/K \gg 1$ and backflow effects can be ignored. The velocity field throughout the sample is then given by Eq. \eqref{eq:simple_shear}, with $\deps=v_{0}/d$ a constant shear rate. A stationary configuration of the average orientation $\theta$ is then found by solving a simplified version of Eq. \eqref{eq:theta_p} of the form
\begin{equation}\label{eq:theta_channel}
\D\partial_{y}^{2}\theta-\frac{\deps}{2}-|\Hp|\sin(p\theta-\Arg\Hp)=0\;,
\end{equation}
with $\theta=\theta(y)$ by virtue of the translational invariance along the $x-$direction imposed by the channel geometry, with boundary conditions
\begin{equation}\label{eq:channel_boundary_conditions}
\theta(0) = \theta_{0}\;,\qquad \theta(d) = \theta_{0}+\Delta\theta\;,	
\end{equation}
with $\Delta\theta$ a constant angle. Before considering the case of general $p$, we review the phenomenon of flow alignment in nematics. In this case, Eq. \eqref{eq:theta_channel} reduces to
\begin{equation}\label{eq:leslie}
\D\partial_{y}^{2}\theta-\frac{\deps}{2}\,\left(1-\frac{\lambda_{2}}{\OPeq}\cos 2\theta\right)=0\;.
\end{equation}
Thus, away of the channel walls, the nematic director orients at an angle
\begin{equation}\label{eq:nematic_flow_alignment}
\theta = \frac{1}{2}\,\arccos\left(\frac{\OPeq}{\lambda_{2}}\right)\;,
\end{equation}
also known as Leslie's angle, with respect to the flow direction \cite{DeGennes:1993}. The latter result applies exclusively to so called {\em flow-aligning} nematics, for which $\lambda_{2}/\OPeq \ge 1$. Nematic liquid crystals with $\lambda_{2}/\OPeq<1$ are known as {\em flow-tumbling} and, when subject to shear, form more complex textures featuring multiple stationary or time-dependent rotations of the nematic director. Near the boundaries, the local orientation $\theta$ inevitably deviates from Leslie's angle in order to match the anchoring direction, as required by Eq. \eqref{eq:channel_boundary_conditions}, thereby creating a boundary layer whose width is approximatively given by $\ls$ in Eq. \eqref{eq:ls}.

Similarly, for $p=4,\,6,\,8\ldots$, Eq. \eqref{eq:theta_channel} reduces to 
\begin{equation}\label{eq:shear_flow_even}
\D\partial_{y}^{2}\theta - \frac{\dot{\epsilon}}{2}\left[1+\left(\frac{\deps}{\deps_{\rm c}}\right)^{p/2-1}\sin p\left(\theta-\frac{\pi}{4}\right)\right]=0\;,
\end{equation}
with $\deps_{\rm c}$ a constant shear rate given by
\begin{equation}\label{eq:critical_shear_rate}
\deps_{\rm c} = 2\left(\frac{p\OPeq}{2\nu_{p}}\right)^{\frac{1}{p/2-1}}\;. 
\end{equation}
Thus, unlike in nematics, the fluid can be either flow-tumbling, for $\deps<\deps_{\rm c}$, or flow-aligning, for $\deps>\deps_{\rm c}$. In the latter case, the director aligns at an angle that progressively approaches the asymptotic value
\begin{equation}\label{eq:theta_infinity}
\theta_{p} = \left(\frac{\pi}{4}+\frac{k\pi}{p}\right)\;{\rm mod}\;\frac{2\pi}{p}\;,
\qquad
k \in \mathbb{Z}\;,
\end{equation}
as $\dot{\epsilon}$ is increased. The integer $k$ depends on the anchoring of the $p-$atic director and can be selected in such a way to minimize the energetic cost of the boundary layer in proximity of the channel walls. Taking, for instance, $\theta_{0}=\Delta\theta=0$, this yields: $\theta_{4}=\pm \pi/4$, $\theta_{6}=\pm \pi/12$, $\theta_{8}=\pm \pi/8$ etc. with the sign is given by $-\sign\dot{\epsilon}$. 

Fig. \ref{fig:5}a show the configurations obtained from a numerical solution of Eq. \eqref{eq:theta_channel} for $3 \le p \le 6$ and different two different Ericsken number values, with boundary condition $\theta_{0}=\Delta\theta=0$. In the case of channel flow, flow alignment is prominent in both tetratics ($p=4$) and hexatics ($p=6$), where the director orientation $\theta$ is uniform in the bulk of the channel and abruptly rotates in proximity of the boundary to comply with the anchoring conditions. The analysis presented here assumes the parameter $\nu_{p}$ constant, but, as anticipated in Sec. \ref{sec:hydrodynamic_equations} [see Eq. \eqref{eq:lambda_nu_effective} in particular], both $\lambda_{p}$ and $\nu_{p}$ could, in principle, depend upon the shear rate $\deps$, as no symmetry prevents this. Whereas at small shear rates these higher order terms would be negligible, the same argument could not be applied in the present context, as the flow alignment phenomenon entailed in Eq. \eqref{eq:shear_flow_even}, holds exclusively at large shear rates. In fact, in the absence of microscopic arguments, one could expect these higher order terms to become comparable to those in Eq. \eqref{eq:shear_flow_even} precisely at $\deps>\deps_{\rm c}$. Hence, in general, we expect both $\deps_{\rm c}$ and the asymptotic flow alignment angle $\theta_{p}$ to be non-universal. In spite of these caveats, the truncated model presented here demonstrates the existence of a region of parameter space, corresponding to $\deps\approx\deps_{\rm c}$, where the terms proportional to $\deps^{\,p/2}$ dominates over all possible higher order terms and flow alignment occurs for arbitary even $p$ values subject to channel confinement. In practice, the occurrence of flow alignment in experiments on driven $p-$atic liquid crystals ultimately depends on the specific material properties of the system, hence on the magnitude of the higher order terms. This situation, however, is no worse than in nematics, where, consistently with Eq. \eqref{eq:nematic_flow_alignment}, the occurrence of flow alignment crucially relies on the specific value of the parameter $\lambda_{2}$.

\begin{figure}
\centering
\includegraphics[width=\columnwidth]{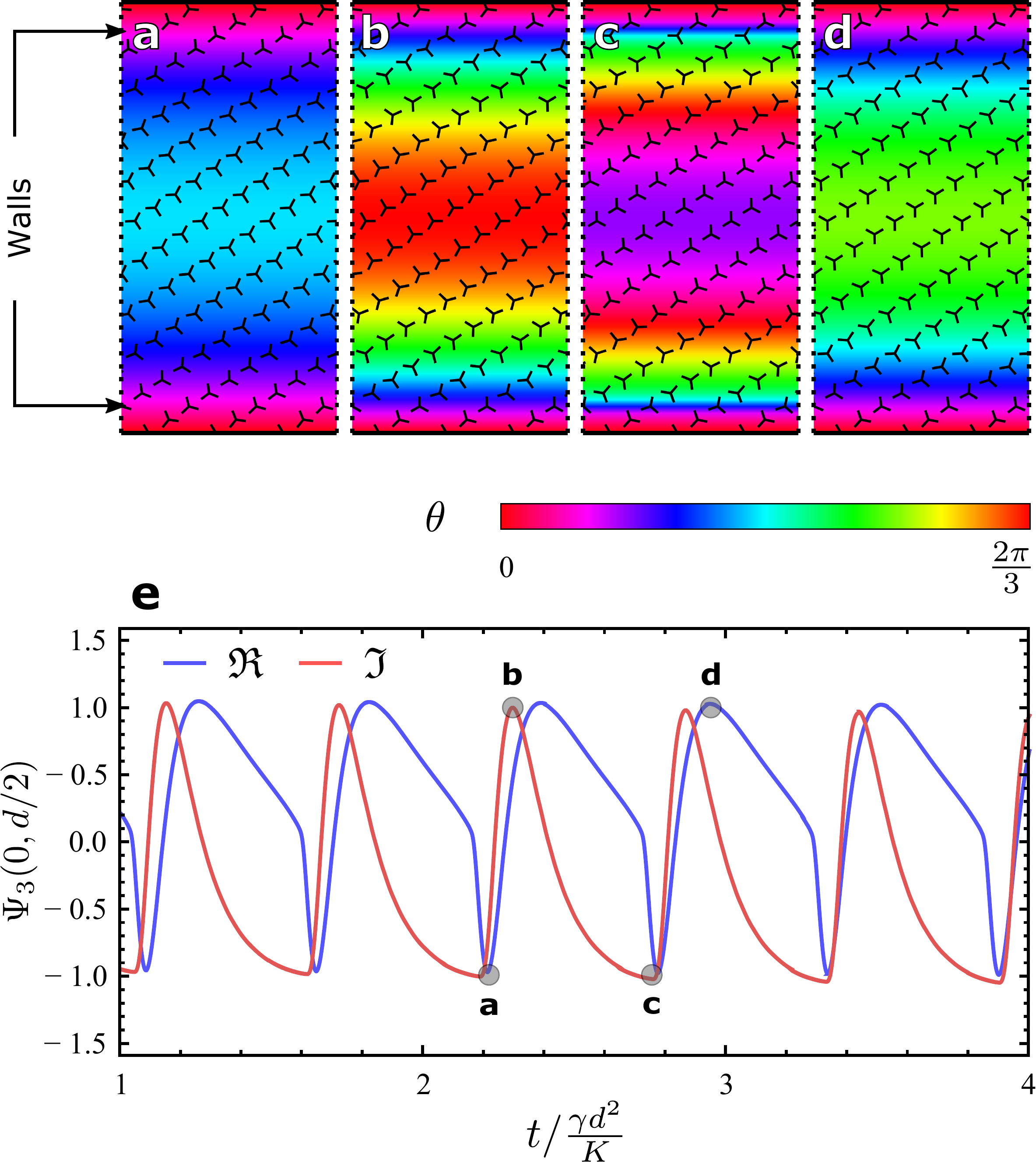}
\caption{\label{fig:6}Examples of flow tumbling in triatic liquid crystals under channel confinment. (a)-(d) Configurations of the triatic director along one tumbling period obtained from a numerical integration of Eqs. (\ref{eq:generic_hydrodynamics}b) and \eqref{eq:psi_1}, slightly above the instability. As in Fig. \ref{fig:5}, the solid horizontal lines denote the channel walls, whereas the dotted vertical lines mark the position of the channel inlet/outlet. The arrows indicate the locations of the ``walls'', where the order parameter periodically vanishes, thus allowing the director in the bulk to temporarily disengage from the boundary. (e) Time-plot of the real (blue) and imaginary (red) parts of the complex order parameter $\Psi_{3}$ at the center of the channel, i.e. $x=0$ and $y=d/2$. The time points corresponding to panels (a)-(d) are marked in the figure. In all panels the parameter values are  $d/\xi_{m}=50$, $\tau_{\rm p}/\tau_{\rm a}=1$, $\gamma/\eta=1$, $\lambda_{3}/d=0.2$, $\OPeq=1$ and the Ericksen number is $\er=35$, thus slightly above the critical Ericksen number $\er_{\rm c}=100/3$.}	
\end{figure}

For $p=3,\,5,\,7\ldots$, on the other hand, $\Hp=0$ because of the uniform shear rate and Eq. \eqref{eq:theta_channel} further simplifies to
\begin{equation}\label{eq:shear_flow_odd}
\D\partial_{y}^{2}\theta-\frac{\deps}{2}=0\;,
\end{equation}
whose solution with the boundary conditions given by Eq. \eqref{eq:channel_boundary_conditions} is
\begin{equation}\label{eq:theta_odd}
\theta(y) = \theta_{0}+\Delta\theta\,\frac{y}{d}+\frac{y(y-d)}{4\ls^{2}}\;.
\end{equation}
Thus for odd $p \ne 1$, the director rotates in such a way to accommodate the vorticity of the imposed shear flow, but without aligning at a specific angle, as can be seen in Fig. \ref{fig:5}b in the case of triatics ($p=3$) and pentatics ($p=5$). As in nematic liquid crystals, however, the stationary configuration described by Eq. \eqref{eq:theta_odd} is unstable to tumbling for finite values of the length scale $\xi_{\rm m}$ defined in Eq. \eqref{eq:mean_field_correlation_length}. In two-dimensional nematics, such an instability takes place via the formation of ``walls'', that is, singular lines located in proximity of the boundaries where the director is highly distorted and the scalar order parameter vanishes \cite{Thampi:2015}. The periodic appearance of walls allows the director in the bulk to temporarily disengage from the boundary and precess at roughly constant angular velocity $\omega_{xy}=-\deps/2$. Fig. \ref{fig:6}, displays the typical tumbling dynamics obtained from a numerical integration of Eqs. (\ref{eq:generic_hydrodynamics}b) and \eqref{eq:psi_1} in the case $p=3$. 

To gain further insight into this instability, we assume the length scale $\xi_{\rm m}$, defined in Eq. \eqref{eq:mean_field_correlation_length}, to be finite and split Eq. \eqref{eq:psi_1} into two coupled partial differential equations for the magnitude $\OP$ and the phase $\theta$ of the complex order parameter $\Psi_{p}$. Using Eq. \eqref{eq:bare_order_parameter}, this gives, after standard algebraic manipulations
\begin{subequations}\label{eq:psi_theta}
\begin{gather}
\D^{-1}\partial_{t}\OP = \nabla^{2}\OP + \frac{\OP}{\xi_{\rm m}^{2}}\left(1-\frac{\OP^{2}}{\OPeq^{2}}-p^{2}\xi_{\rm m}^{2}|\nabla\theta|^{2}\right)\;,\\
\D^{-1}\OP\partial_{t}\theta = \OP\left(\nabla^{2}\theta-\frac{1}{2\ls^{2}}\right)+2\nabla\OP\cdot\nabla\theta\;,
\end{gather}
\end{subequations}
where we used again Eq. \eqref{eq:simple_shear} to express the velocity $\bm{v}$ and its derivatives in terms of the shear rate $\deps$. The last three terms on the right-hand side of Eq. (\ref{eq:psi_theta}a) set the magnitude of the scalar order parameter $\OP$, which, in turn, is positive by construction and vanishes in the isotropic phase: i.e. $\OP \ge 0$. At low shear rates, the latter condition can be fulfilled throughout the entire channel and the solution of Eqs. \eqref{eq:psi_theta} is given, at the quadratic order in $d/\ls$, by $\OP=\OPeq$ and Eq. \eqref{eq:theta_odd}. As the shear rate is increased, the distortion of the $p-$atic director is initially compensated by a decrease of the scalar order parameter, until, for high shear rates, this becomes virtually negative, thereby violating the positivity requirement. As the $p-$atic director is more highly distorted near the boundaries of the channel, the critical shear rate associated with the tumbling instability can be found by demanding
\begin{equation}\label{eq:tumbling_instability}
\OP^{2} = \OPeq^{2}\left(1-p^{2}\xi_{\rm m}^{2}|\nabla\theta|^{2}\right) \ge 0\;,
\end{equation}	
at $y=0$ and $y=d$. Next, assuming the anchoring conditions to be the same on both boundaries (i.e. $\Delta\theta=0$) and using Eq. \eqref{eq:theta_odd} to express $|\nabla\theta|_{y=0,\,d}=d/(2\ls^{2})$, solving Eq. \eqref{eq:tumbling_instability} readily yields the following stability criterion for the static configuration:
\begin{equation}\label{eq:positivity_condition}
\frac{\xi_{\rm m}d}{\ls^{2}} \ge \frac{2}{p}\;,	
\end{equation}
from which one finds the critical Ericksen number associated with the tumbling transition in the form
\begin{equation}\label{eq:critical_er}
\er_{\rm c} = \frac{(\eta/\gamma)(d/\xi_{\rm m})}{p/2}\;,
\end{equation}	
in perfect agreement with our numerical solutions of Eqs. (\ref{eq:generic_hydrodynamics}b) and \eqref{eq:psi_1}, which additionally include backflow effects (Fig. \ref{fig:6}). 

Some comments are in order. Eqs. \eqref{eq:positivity_condition} and \eqref{eq:critical_er} hold exclusively for odd $p$ values, whereas for even $p$ values flow alignment prevents the tumbling instability from taking place. The critical Ericksen number is a monotonically decreasing function of $p$ and vanishes in the limit $p \rightarrow \infty$, when isotropy is restored at the microscopic scale. Furthermore, since $\eta\approx\gamma$ in most liquid crystals \cite{Kleman:2003} and $\xi_{\rm m}$ has the same order of magnitude of the size of the microscopic building blocks, we expect that the tumbling instability discussed here is accessible in experiments on colloidal $p-$atics (see e.g. Fig. \ref{fig:1}). Finally, this instability shares some resemblances with the Silsbee criterion in superconducting wires (see e.g. Ref.~\cite{Tinkham:1975}). 

\subsection{\label{sec:taylor_couette}Taylor-Couette flow}

As a second example of flow alignment in $p-$atics, we consider the two-dimensional analog of Taylor-Couette flow, that is, the flow induced inside an annulus delimited by two concentric circles of radii $R_{1}$ and $R_{2}=R_{1}+d$, with $d$ the width of the annulus, rotating at angular velocities $\Omega_{1}$ and $\Omega_{2}$ respectively. At large shear rate, where backflow effects are negligible, the velocity field becomes identical to that of an isotropic fluid, given by \cite{Taylor:1923}
\begin{equation} \label{eq:taylor_couette_velocity}
\bm{v} = \left(Ar+\frac{B}{r}\right)\bm{e}_{\phi}\;,
\end{equation}
where $r=\sqrt{x^{2}+y^{2}}$ is the distance from the center of the annulus, $\bm{e}_{\phi}=-\sin\phi\,\bm{e}_{x}+\cos\phi\,\bm{e}_{r}$ and $\bm{e}_{r}=\cos\phi\,\bm{e}_{x}+\sin\phi\,\bm{e}_{y}$, with $\phi=\arctan y/x$, orthonormal basis vectors in the longitudinal and transverse direction respectively and we have set
\begin{equation}
A = \frac{R_{2}^{2}\Omega_{2}-R_{1}^{2}\Omega_{1}}{R_{2}^{2}-R_{1}^{2}},\qquad
B = \frac{R_{1}^{2}R_{2}^{2}(\Omega_{1}-\Omega_{2})}{R_{2}^{2}-R_{1}^{2}}\;,
\end{equation}
from which the components of the strain rate and vorticity tensor can be readily computed in the form
\begin{subequations}\label{eq:taylor_couette_strain_rate}
\begin{gather}
u_{rr} = u_{\phi\phi} = 0\;,\\[4pt]
u_{r\phi} = u_{\phi r} = -\frac{B}{r^{2}}\;,\\[5pt]
\omega_{r\phi} = -  \omega_{\phi r} = A\;.
\end{gather}
\end{subequations}
Thus, unlike in the case of simple shear flow discussed in Sec. \ref{sec:simple_shear}, the strain rate across the Taylor-Couette cell and both the linear and non-linear terms in the flow alignment field $\Hp$ do not vanish identically. The typical strain rate of the flow is given by 
\begin{equation}
\dot{\epsilon} = \frac{\Omega_{2}R_{2}-\Omega_{1}R_{1}}{R_{2}-R_{1}}\;.
\end{equation}
Eq. \eqref{eq:theta_p} can be expressed in polar coordinates by setting $\alpha=\theta-\phi$. Thus, using Eqs. \eqref{eq:taylor_couette_velocity} and \eqref{eq:taylor_couette_strain_rate} and assuming $\partial_{\phi}\alpha=0$ by virtue of the rotational symmetry of the annulus, yields the following equation for a stationary configuration of the average orientation $\alpha$:
\begin{equation}\label{eq:alpha}
\D\left(\partial_{r}^{2}\alpha+\frac{1}{r}\,\partial_{r}\alpha\right) 
- \frac{B}{r^{2}}
- |\Hp|\sin\left(p\alpha-\Arg\Hp\right) = 0 \;,
\end{equation}
where the flow alignment field now takes the form:
\begin{equation}
\mathfrak{H}_{p} = \frac{\hp}{r^{p}}\;,	
\end{equation}
with $\hp$ a complex number given by
\begin{multline}
\chi = \frac{2}{p\OPeq}
\Big\{
-i \lambda_{p} B(-\sqrt{2})^{p-2}(p-1)!\\
+\nu_{p} B^{\lfloor p/2 \rfloor}[-\sqrt{2}\,(p-1)]^{p\,{\rm mod}\,2}e^{-i\lfloor p/2 \rfloor\frac{\pi}{2}}
\Big\}\;.
\end{multline}
Now, in nematics, $\chi=-i\lambda_{2}B/\OPeq$ and Eq. \eqref{eq:alpha} yields again Leslie's angle in the rotating frame $\{\bm{e}_{r},\bm{e}_{\phi}
\}$,  :
\begin{equation}
\alpha = \frac{1}{2}\arccos\left(-\frac{\OPeq}{\lambda_{2}}\right)\;.
\end{equation}
By contrast, for any $p\ge 3$ value, the $r-$dependence of the right-hand side of Eq. \eqref{eq:alpha} does not cancel. In this case, defining
\begin{equation}
\Rs=\left|\frac{\hp}{B}\right|^{\frac{1}{p-2}}\;,
\end{equation}
the stationary configuration of the local orientation $\alpha$ is found to obey the ordinary differential equation
\begin{multline}\label{eq:alpha_steady}
\D\left(\partial_{r}^{2}\alpha+\frac{1}{r}\,\partial_{r}\alpha\right) \\
- \frac{B}{r^{2}}\left[1+\left(\frac{\Rs}{r}\right)^{p-2}\sin\left(p\alpha-\Arg\Hp\right)\right] = 0\;,
\end{multline}
with and boundary conditions
\begin{equation}
\alpha(R_{1}) = \alpha_{0}\;,\qquad
\alpha(R_{2}) = \alpha_{0}+\Delta\alpha\;.
\end{equation}
In the absence of flow alignment effects, $\Rs=0$ and the solution of Eq. \eqref{eq:alpha_steady} is given by
\begin{equation}\label{eq:alpha_no_flow_alignment}
\alpha 
= \alpha_{0} 
+ \frac{\Delta\alpha}{\log R_{2}/R_{1}}\,\log \frac{r}{R_{1}}
+ \frac{B}{2\D}\log\left(\frac{r}{R_{1}}\right)\log\left(\frac{r}{R_{2}}\right)\;,
\end{equation}
and the $p-$atic director tumbles across the Taylor-Couette cells depending on the ratio $B/\D\sim (R_{1}/\ell_{\rm s})^{2}$, with $\ell_{\rm s}$ the shear length scale given in Eq. \eqref{eq:ls}.

Conversely, for non-vanishing $\lambda_{p}$ and $\nu_{p}$ values, and with only exception for $p=3$, increasing the shear rate $\deps$ results in an increase of the length scale $\Rs$ until, for $\Rs \gg r$, flow alignment effects becomes dominant and the $p-$atic director uniformly aligns at an angle
\begin{equation}\label{eq:taylor_couette_flow_alignment}
\alpha_{p} = \left(\lim_{\dot{\epsilon}\rightarrow\infty}\frac{\Arg\Hp}{p}+\frac{k\pi}{p}\right)\;{\rm mod}\;\frac{2\pi}{p}\;,
\qquad
k \in \mathbb{Z}\;,
\end{equation}
in the bulk of the Taylor-Coutte cell. The integer $k$ depends again on the anchoring conditions and is chosen in such a way to minimize the energetic cost of the boundary layer near the edges. As in the large $\dot{\epsilon}$ limit, the value of $\Arg\Hp$ converges towards either $0$, $\pm \pi/2$ and $\pm \pi$, depending on the sign of the constants $\lambda_{p}$, $\nu_{p}$ and $B$. Thus, the asymptotic bulk orientation $\alpha_{p}$ is non-universal and, unlike in the case of simple shear flow discussed in Sec. \ref{sec:simple_shear}, can be used in order to infer information about the material parameters.  

Fig. \ref{fig:5}b some examples of Taylor-Couette flow in $p-$atics with $3\le p \le 6$, obtained from a numerical integration of Eq. \eqref{eq:alpha_steady} with boundary conditions $\alpha_{0}=\pi/2$ and $\Delta\alpha=0$ and the same $\er$ values already considered in the case of channel flow. Unlike the latter, here the spatial dependence of the strain rate $u_{r\phi}$ and the vorticity $\omega_{r\phi}$ render the flow alignment field $\Hp$ non-vanishing regardless of the specific $p$ value and the director is always found to flow align at large shear rates. 

\begin{figure}[t]
\centering
\includegraphics[width=\columnwidth]{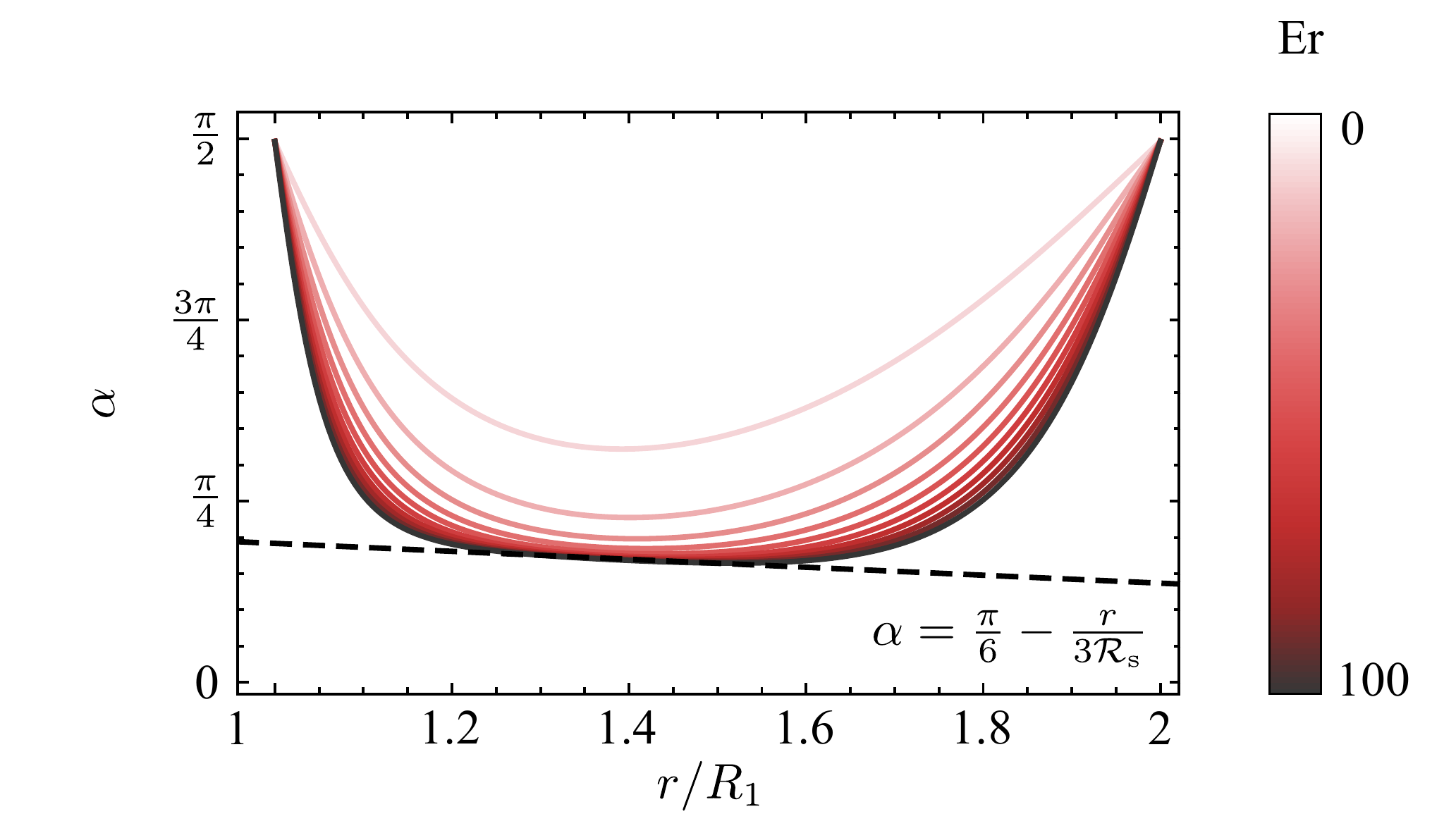}
\caption{\label{fig:triatic_flow_alignment}Numerical solution of Eq. \eqref{eq:alpha_steady} with boundary conditions $\alpha_{0}=\pi/2$ and $\Delta\alpha=0$ for ten different values of the Ericsken number. Unlike for $p>3$, the bulk orientation of the triatic director in a Taylor-Couette flow do not flow align at angle $\alpha_{p}$ given in Eq. \eqref{eq:taylor_couette_flow_alignment}, but, for large $\er$ values, it approaches the linearly decreasing function given in Eq. \eqref{eq:alpha_approx} and whose slope depends uniquely on the triatic flow alignment parameter $\lambda_{3}$. The parameter values are $\tau_{\rm p}/\tau_{\rm a}=1$, $\lambda_{3}/d=1.5$, $R_{1}/d=1$, $R_{2}/d=2$ and $\Omega_{1}=0$.}
\end{figure}

For $p=3$, $\chi=4i\sqrt{2}\lambda_{3}B/(3\OPeq)$ and the length scale $\Rs$ does not diverge for large shear rates. In this case, the local orientation $\alpha$ does not approach the asymptotic value given by Eq. \eqref{eq:taylor_couette_flow_alignment}, but varies along the radial direction similarly to the statically tumbling configuration described by Eq. \eqref{eq:theta_odd}. Yet, for large shear rate, thus large $B$ values, the last two terms on the left-hand side of Eq. \eqref{eq:alpha_steady}, expressing the hydrodynamic torque experienced by the $p-$atic director, overweight the restoring torques originating from the entropic elasticity of the triatic phase. Away from the edges of the annulus, the configuration of the position-dependent orientation $\alpha$ can be found from the solution of the trigonometric equation
\begin{equation}\label{eq:triatic_flow_alignment}
\cos 3\alpha = -\frac{r}{\Rs}\;.
\end{equation}
Since $R_{1} \le r \le R_{2}$, this equation admits a real-valued solution across the entire annulus only if $\Rs>R_{2}$. In particular, when $\Rs\gg R_{2}$, expanding the left-hand side of Eq. \eqref{eq:triatic_flow_alignment} about $\alpha=\pi/6$ yields
\begin{equation}\label{eq:alpha_approx}
\alpha \approx \frac{\pi}{6}-\frac{r}{3\Rs}\;,	
\end{equation}
in good agreement with the numerical solution of Eq. \eqref{eq:alpha_steady} at high shear rates (Fig. \ref{fig:triatic_flow_alignment}). Since $\Rs$ is a simple linear function of triatic flow alignment parameter $\lambda_{3}$, i.e. $\Rs=(4\sqrt{2}/3)\lambda_{3}$, Eq. \eqref{eq:alpha_approx} provides a potentially simple and efficient strategy to measure $\lambda_{3}$ in experiments on colloidal triatics, like those shown in Fig. \ref{fig:1}a.

\section{\label{sec:conclusions}Conclusions}

In this article we developed a comprehensive hydrodynamic theory of $p-$atic liquid crystals in two dimensions. These phases of matter naturally arise in suspensions of $p-$fold symmetric colloidal particles at an interface (Fig. \ref{fig:1}) and in two-dimensional melting, as an intermediate state between crystalline solids and isotropic liquids. In the realm of biological matter, nematic order (i.e. $p=2$) is commonly found at both cellular and sub-cellular scale. Micro-colonies of rod-shaped sessile bacteria \cite{You:2018,DellArciprete:2018,Li:2019}, monolayer of motile cells \cite{Duclos:2014,Kawaguchi:2017,Duclos:2017}, in vitro mixtures of cytoskeletal filaments and motor proteins \cite{Asano:2009,Sanchez:2012,DeCamp:2015,Lemma:2019}, are prominent examples of biological nematic fluids in two dimensions. These examples have recently attracted the attention of a large and multidisciplinary community at the crossroads between soft matter and biophysics. Perhaps more remarkably, recent computational work by Li and Pica Ciamarra has suggested that hexatic order (i.e. $p=6$) could exist in confluent epithelial tissues and cell layers \cite{Li:2018}. As in the case of two-dimensional melting, these biological liquid crystals occupy a region of phase-space intermediate between solid and liquid, but, unlike in the classic KTHNY scenario, the transition is mainly driven by the cells' geometrical frustration and carries over even in the absence of thermal fluctuations.

Previous hydrodynamic theories of $p-$atics \cite{Zippelius:1980a,Zippelius:1980b} have ${\rm O}(2)$ rotational symmetry \cite{Zippelius:1980a,Zippelius:1980b}, which is higher than the actual $p-$fold symmetry of $p-$atic phases. In this paper, we have gone beyond this picture, using a phenomenological approach involving the $p-$atic tensor order parameter $\bm{Q}_{p}$, whose algebraic structure directly embodies the discrete rotational symmetry of $p-$atics. We identified additional couplings between $p-$atic order and flow which break the ${\rm O}(2)$ rotational symmetry of earlier models down to the $p-$fold rotational symmetry of $p-$atics. These are linear and non-linear functions of the strain-rate and $p-$atic tensor order parameter $\bm{Q}_{p}$. These novel couplings leave a distinct signature on the high shear rate dynamics, which may cause  the $p-$atic director to align at specific system-dependent angles with respect to the underlying velocity field at sufficiently high shear rates. Unlike in three-dimensional nematics, in which bulk flow alignment occurs independent of the applied shear rate, the nonlinear nature of the coupling between orientation and flow for $p>2$ renders this phenomenon shear rate dependent, so that flow alignment can only occur at large shear rates. 

Our approach is also particularly well suited for numerical simulations of coarsening phenomena and other processes characterized by the occurence of topological defects. Unlike the local orientation $\theta$, the order parameter tensor $\bm{Q}_{p}$ is everywhere defined, including within the core of disclinations of arbitrary winding number. This allows for an efficient description of the dynamics of defective configurations, especially in the absence of regular patterns, for which the location of these orientational singularities cannot be predicted a priori.

Finally, using fluctuating hydrodynamics, we have demonstrated that a shear flow of arbitrary finite shear rate has the remarkable effect of turning quasi-long-ranged orientational order, i.e. the hallmark of two-dimensional liquid crystals at equilibrium, into long-ranged order. We have also shown that fluctuation effects prevent flow alignment at low shear rates for {\em any} value of $p$, even $p=1$ and $p=2$, for which mean field theory would predict flow alignment at arbitrarily small shear rates.

\acknowledgements

We are indebted with Massimo Pica Ciamarra for insightful discussions. This work is partially supported by the ERC-CoG grant HexaTissue (L.G.) and by Netherlands Organization for Scientific Research (NWO/OCW), as part of the Vidi scheme (N.S. and L.G.) and the Frontiers of Nanoscience program (L.G.). JT thanks the Max-Planck Institut f\"ur Physik Komplexer Systeme,  Dresden, Germany for their hospitality, and their support through the Martin Gutzwiller Fellowship, and the Lorentz Center of the University of Leiden, Leiden, NL, for their support during a brief visit there,  while a portion of this work was underway.

\appendix

\section{\label{app:delta_tensor}Derivation of Eq. \eqref{eq:delta_tensor}}

Following Ref. \cite{Hess:2015}, in this Appendix, we provide a derivation of Eq. \eqref{eq:delta_tensor} in terms of derivatives of multipole potentials. Multipole potentials are tensorial solutions of the Laplace equation. These can be hierarchically obtained starting from the Green function $X_{0}= \log |\bm{r}|/d$, with $d$ an arbitrary length scale. For $|\bm{r}|>0$, both $X_{0}$ and its derivatives $\nabla^{\otimes p}X_{0}=\partial^{p}_{i_{1}i_{2}\cdots\,i_{p}}X_{0}$ are solutions of the Laplace equation, from which one can define the family of tensorial solutions:
\begin{equation}
\bm{X}_{p} = -\nabla\bm{X}_{p-1} = (-1)^{q}\nabla^{\otimes q}\bm{X}_{p-q}\;,	
\end{equation}
with $q\in\mathbb{N}$. The tensorial functions $\bm{X}_{p}$ are proportional to the irreducible tensor $\traceless{\bm{r}^{\otimes p}}$ constructed from the components of the position vector $\bm{r}$, which, in turn, are related to $\bm{\Delta}_{p,p}$ \cite{Hess:2015}, as follows: 
\begin{equation}\label{eq:r_vs_delta}
\partial_{i_{1}i_{2}\cdots\,i_{p}}^{p}\traceless{r_{j_{1}}r_{j_{2}}\cdots\,r_{j_{p}}} = p!\,\Delta_{i_{1}i_{2}\cdots\,i_{p}j_{1}j_{2}\cdots\,j_{p}}\;.	
\end{equation}
Now, starting from $X_{0}$ and using an inductive construction, one can show that:
\begin{equation}\label{eq:r_vs_x}
\traceless{\bm{r}^{\otimes p}} = - \frac{1}{[2(p-1)]!!}\,r^{2p}\bm{X}_{p}\;.
\end{equation}
Finally, combining Eqs. \eqref{eq:r_vs_delta} and \eqref{eq:r_vs_x} yields Eq. \eqref{eq:delta_tensor}.

\section{\label{app:stress}Calculation of the stresses}

\subsection{Static stress}

In order to calculate the elastic stress arising in the system in response to a static deformation of the $p-$atic tensor, let us consider an arbitrarily small virtual displacement of the form $\bm{r} \rightarrow \bm{r}+\delta\bm{r}$ acting upon a fluid patch $\Omega$. The associated free energy variation is given by:
\begin{equation}
\delta F = \int_{\Omega}{\rm d}A\, \delta f + \oint_{\partial\Omega} {\rm d}\ell\,f\,\delta\bm{r}\cdot\bm{N}\;,
\end{equation}
where $f$ is the free energy density, and the second term accounts for the free energy change associated with a displacement of the boundaries of the patch, whose outward-pointing normal vector is indicated with $\bm{N}$. Now, defining $\delta\bm{Q}_{p}$ as the change in the $p-$atic tensor induced by the virtual displacement, and expanding $f$ at the linear order in $\delta\bm{Q}_{p}$, yields:
\begin{multline}
\delta F = 
- \int_{\Omega} {\rm d}A\,\bm{H}_{p}\odot\delta\bm{Q}_{p} \notag \\
+ \oint_{\partial \Omega} {\rm d}\ell N_{j}\left[f\delta r_{j}+\frac{\partial f}{\partial(\partial_{j}\bm{Q}_{p})}\odot\delta\bm{Q}_{p}\right]\;,
\end{multline}
where:
\begin{equation}
\bm{H}_{p} 
= - \frac{\delta F}{\delta \bm{Q}_{p}}
= - \frac{\partial f}{\partial \bm{Q}_{p}}+\partial_{i}\left[\frac{\partial f}{\partial(\partial_{i}\bm{Q}_{p})}\right]\;,
\end{equation}
is the molecular tensor. Next, performing a gradient expansion of $\delta\bm{Q}_{p}$ by writing $\delta\bm{Q}_{p}=-\delta r_{i}\partial_{i}\bm{Q}_{p}$ yields, after standard algebraic manipulations:
\begin{multline}\label{eq:free_energy_varation}
\delta F = 
- \int_{\Omega} {\rm d}A\,\bm{Q}_{p}\odot\partial_{i}\bm{H}_{p}\,\delta r_{i} \\
+ \oint_{\partial\Omega} {\rm d}\ell\,N_{j}\left[\left(f+\bm{H}_{p}\odot\bm{Q}_{p}\right)\delta_{ij}-\frac{\partial f}{\partial(\partial_{j}\bm{Q}_{p})}\odot\partial_{i}\bm{Q}_{p}\right]\delta r_{i}\;.
\end{multline} 
The mechanical work performed on the fluid patch, the other hand, can be expressed as:
\begin{equation}\label{eq:mechanical_work}
W 
= \hspace{-0.5ex}\int_{\Omega} {\rm d}A\,\sigma_{ij}^{({\rm e})}\epsilon_{ji} 
=-\hspace{-0.5ex}\int_{\Omega} {\rm d}A\,\partial_{j}\sigma_{ij}^{({\rm e})}\,\delta r_{j} 
+ \oint_{\partial\Omega} {\rm d}\ell\,\sigma_{ij}^{{(\rm e)}}\delta r_{i}N_{j},
\end{equation}
with $\epsilon_{ji}=\partial_{j}\delta r_{i}$ the strain tensor. Comparing the boundary terms in Eqs. \eqref{eq:free_energy_varation} and \eqref{eq:mechanical_work} allows us to identify the elastic stress:
\begin{equation}
\sigma_{ij}^{({\rm e})} 
= \left(f+\bm{H}_{p}\odot\bm{Q}_{p}\right)\delta_{ij}
- \frac{\partial f}{\partial(\partial_{j}\bm{Q}_{p})}\odot\partial_{i}\bm{Q}_{p}\;,
\end{equation}
whereas a comparison of the bulk integrals yields a Gibbs-Duhem equation for $p-$atics:
\begin{equation}
\partial_{j}\sigma^{({\rm e})}_{ij} = \bm{Q}_{p}\odot\partial_{i}\bm{H}_{p}\;.	
\end{equation}
The procedure outlined above assumes that the mapping $\bm{r}\rightarrow\bm{r}+\delta\bm{r}$ leaves the area of the fluid patch unchanged, hence $\tr\bm{\epsilon}=\nabla\cdot\delta\bm{r}=0$. This constrained could be explicitly accounted for in the calculation of the free energy variation by considering the alternative functional $F'=F+\int {\rm d}A\, \mu \nabla\cdot\delta\bm{r}$, with $\mu=\mu(\bm{r})$ a Lagrange multiplier. This, however, leads to an additional isotropic term that can be incorporated into the pressure. 

Finally, taking $f$ as given in Eq. \eqref{eq:free_energy} yields, up to the aforementioned terms that can be incorporated into the isotropic pressure, Eq. \eqref{eq:elastic_stress}.

\subsection{Dynamic stress}

In order to calculate the reactive stresses arising from linear couplings between the velocity gradient $\nabla\bm{v}$ and the $\bm{Q}_{p}$ tensor, one needs to cast the entropy production rate density, Eq. \eqref{eq:entropy_production_1}, in the form:
\begin{multline}
\sigma_{ij}^{(\rm v)}\partial_{j}v_{i} + \bm{H}_{p}\odot\frac{D\bm{Q}_{p}}{Dt} \\
= \left( \sigma_{ij}+P\delta_{ij}-\sigma_{ij}^{({\rm e})}-\sigma_{ij}^{({\rm d})}\right) \partial_{j}v_{i} + \nabla\cdot(\cdots)\;, 
\end{multline}
from which one readily obtains Eq. \eqref{eq:reactive_stress}, hence $\bm{\sigma}^{({\rm d})}$, by recognizing that $\bm{\sigma}=\bm{\sigma}^{{(\rm r})}$ when $T\dot{S}=0$. The problem of computing the dynamic contribution to the reactive stress is then reduced to the simple task of expressing the inner product between the molecular tensor and the material derivative of the $p-$atic tensor in the form $\sigma_{ij}^{({\rm d})}\partial_{j}v_{i}$, up to boundary terms. 

In order to perform this computation, we ignore the nonlinear term $\bm{N}_{p}$ and write:
\begin{align}\label{eq:stress_lambda}
\bm{H}_{p}\odot\frac{D\bm{Q}_{p}}{Dt} = I_{1}+I_{2}+\bar{\lambda}_{p}\bm{Q}_{p}\odot\bm{H}_{p}\,\delta_{ij}\partial_{j}v_{i}\;.
\end{align}
The inner product $I_{1}$ originates from the corotational derivative $\bm{Q}_{p}$ and is given by
\begin{align}\label{eq:hdotdqdt}
I_{1}
&= p \bm{H}_{p}\odot \traceless{\bm{Q}_{p}\cdot\bm{\omega}} \phantom{\frac{1}{2}} \notag \\
&=-\frac{p}{2}\left(H_{k_{1}k_{2}\cdots\,i}Q_{k_{1}k_{2}\cdots\,j}\partial_{i}v_{j}-H_{k_{1}k_{2}\cdots\,i}Q_{k_{1}k_{2}\cdots\,j}\partial_{j}v_{i}\right) \notag \\
&=-\frac{p}{2}\left(Q_{k_{1}k_{2}\cdots\,i}H_{k_{1}k_{2}\cdots\, j}- H_{k_{1}k_{2}\cdots\, i}Q_{k_{1}k_{2}\cdots\,j}\right)\partial_{j}v_{i}\;.
\end{align}
Analogously, the inner product $I_{2}$ can be computed as
\begin{align}\label{eq:hdotl}
I_{2} 
&= \bm{H}_{p}\odot\traceless{\nabla^{\otimes p-2}\bm{u}}\phantom{\frac{1}{2}} \phantom{\frac{1}{2}}\notag \\ 
&= \lambda_{p}H_{k_{1}k_{2}\cdots\,ij}\partial_{k_{1}k_{2}\cdots\,k_{p-2}j}^{p-1}v_{i}  \phantom{\frac{1}{2}} \phantom{\frac{1}{2}}\notag \\
&= \lambda_{p}(-1)^{p-2}\partial_{k_{1}k_{2}\cdots\,k_{p-2}}^{p-2}H_{k_{1}k_{2}\cdots\,ij}\partial_{j}v_{i} + \nabla\cdot(\cdots)\;,\phantom{\frac{1}{2}}
\end{align}
where the second line is derived from the first one by $p-2$ applications of the chain-rule. Combining Eqs. \eqref{eq:hdotdqdt} and \eqref{eq:hdotl} readily yields Eq. \eqref{eq:dynamic_stress}.

\section{\label{app:long_ranged_order}Long-ranged order under shear}

\subsection{Derivation of Eqs. \eqref{eq:theta_simplified_1} and \eqref{eq:noise_correlation}}

In the regime where Eq. \eqref{eq:re_vs_er} holds, inertial and $\mathcal{O}(|\nabla\vartheta|^{2})$ terms can be neglected. Then, approximating 
\begin{equation}
\bm{\sigma}^{({\rm r})} = -P_{0}\mathbb{1}+\frac{K}{2}\,\bm{\varepsilon}\nabla^{2}\vartheta+\mathcal{O}(|\nabla\vartheta|^{2})\;,
\end{equation}
by virtue of Eq. \eqref{eq:theta_stress}, and decomposing the velocity as in Eq. \eqref{eq:velocity_decomposition}, allows us to cast Eqs. \eqref{eq:fluctuating_hydrodynamics} in the simplified form
\begin{subequations}\label{eq:fluctuating_hydrodynamics_real}
\begin{gather}
0 = \eta\nabla^{2}\delta\omega - \frac{K}{2}\,\nabla^{4}\vartheta'+\xi^{(\omega)}\;,\\
(\partial_{t}+\langle \bm{v} \rangle \cdot\nabla)\vartheta' = \D\nabla^{2}\vartheta'+\frac{\delta\omega}{2}+\xi^{(\vartheta)}\;,
\end{gather}	
\end{subequations}
where $\delta\omega=\partial_{x}\delta v_{y}-\partial_{y}\delta v_{x}$ and $\vartheta'=\vartheta+\deps t /2$. Next, dropping the prime for sake of conciseness and expressing Eqs. \eqref{eq:fluctuating_hydrodynamics_real} in Fourier space gives
\begin{subequations}\label{eq:fluctuating_hydrodynamics_fourier}
\begin{gather}
0 = -\eta q^{2}\delta\omega(\bm{q},t)-\frac{K}{2}\,q^{4}\vartheta(\bm{q},t)+\xi^{(\omega)}(\bm{q},t)\;,\\
(\partial_{t}+\langle \bm{v} \rangle \cdot\nabla)\vartheta(\bm{q},t) = -\D q^{2}\vartheta(\bm{q},t)	\notag\\
+\frac{\delta\omega(\bm{q},t)}{2}+\xi^{(\vartheta)}(\bm{q},t)\;,
\end{gather}	
\end{subequations}
where the correlation function of the Fourier amplitudes of the random fields $\xi^{(\omega)}$ and $\xi^{(\vartheta)}$
\begin{multline}\label{eq:noise_correlation_fourier}
\left\langle \xi^{(\alpha)}(\bm{q},t)\xi^{(\beta)}(\bm{q}',t')\right\rangle\\
= 2k_{\rm B}T(2\pi)^{2}\left(\frac{1}{\gamma}\,\delta_{\alpha\vartheta}\delta_{\beta\vartheta}+\eta q^{4}\delta_{\alpha\omega}\delta_{\beta\omega}\right)\delta(\bm{q}+\bm{q}')\delta(t-t')\;.	
\end{multline}
Solving Eq. (\ref{eq:fluctuating_hydrodynamics_fourier}a) with respect to $\delta\omega(\bm{q},t)$ readily yields
\begin{equation}
\delta\omega(\bm{q},t) = -\frac{K}{2\eta}\,q^{2}\vartheta(\bm{q},t)+\frac{\xi^{(\omega)}(\bm{q},t)}{\eta q^{2}}\;.	
\end{equation}
Replacing this in Eq. (\ref{eq:fluctuating_hydrodynamics_fourier}b) then gives
\begin{equation}\label{eq:theta_fourier}
\partial_{t}\vartheta(\bm{q},t) = -q^{2}\De\vartheta(\bm{q},t)+\xi(\bm{q},t)\;,	
\end{equation}
where 
\begin{equation}
\xi(\bm{q},t) = \xi^{(\vartheta)}(\bm{q},t)+\frac{\xi^{(\omega)}(\bm{q},t)}{2\eta q^{2}}\;,	
\end{equation}
is an effective rotational noise, whose correlation function can be readily computed from Eq. \eqref{eq:noise_correlation_fourier}, to give
\begin{equation}\label{eq:effective_noise_correlation_fourier}
\left\langle \xi(\bm{q},t)\xi(\bm{q}',t')\right\rangle
= \frac{2k_{\rm B}T}{\Ge}\,(2\pi)^{2}\delta(\bm{q}+\bm{q}')\delta(t-t')\;,	
\end{equation}
where $\Ge=K/\De$. Finally, expressing Eqs. \eqref{eq:theta_fourier} and \eqref{eq:effective_noise_correlation_fourier} in real space, one obtains Eqs. \eqref{eq:theta_simplified_1} and  \eqref{eq:noise_correlation}.

\subsection{Derivation of Eq. \eqref{eq:g}}

Calculating the $p-$atic correlation function $\langle \psi_{p}^{*}(\bm{r})\psi_{p}(\bm{0}) \rangle$ requires computing the orientational structure factor $\langle|\vartheta(\bm{q},t)|^{2}\rangle$ appearing in Eq. \eqref{eq:q_integral_1}. Following Onuki~\cite{Onuki:1979b} and Ramaswamy~\cite{Ramaswamy:1984}, this can be achieved by solving the stochastic partial differential equation Eq. \eqref{eq:theta_simplified_1}. As in the previous Subsection, we incorporate the vorticity into the definition of the $\vartheta$ field. This yields:
\begin{equation}\label{eq:theta_simplified_1_app}
\partial_{t} \vartheta' + \dot{\epsilon}y\,\partial_{x}\vartheta' = \De \nabla^{2}\vartheta' + \xi\;,
\end{equation}
where we have set again $\vartheta'=\vartheta+\deps t/2$. Notice that, since $\omega_{xy}=-\deps/2$ is uniform in space, this change of variable does not affect the equal time connected correlation function: i.e. $\langle [\vartheta'(\bm{r},t)-\vartheta'(\bm{0},t)]^{2} \rangle = \langle [\vartheta(\bm{r},t)-\vartheta(\bm{0},t)]^{2} \rangle$. 

Next, we can eliminate the convective term in Eq. \eqref{eq:theta_simplified_1_app} by performing the following position-dependent Galilean transformation:
\begin{equation}\label{eq:galilean}
\left\{
\begin{array}{l}
x' = x-v_{x}t = x-\dot{\epsilon}yt\;,\\[3pt]
y' = y\;,\\[5pt]
t' = t\;.
\end{array}
\right.
\end{equation}
This yields:
\begin{equation}\label{eq:theta_simplified_2}
\partial_{t'} \vartheta' = \De\left[\partial_{x'}^{2}+\left(\partial_{y'}-\dot{\epsilon}t\partial_{x'}\right)^{2}\right]\vartheta' + \xi\;,
\end{equation}
or, in Fourier space,
\begin{equation}\label{eq:theta_simplified_3}
\partial_{t} \vartheta = -\mathcal{L}(\bm{q},t)\vartheta + \xi\;,
\end{equation}
where:
\begin{equation}
\mathcal{L}(\bm{q},t) = \De\left[q_{x}^{2}+(q_{y}-\dot{\epsilon}tq_{x})^{2}\right]\;,	
\end{equation}
and we have dropped the prime. The general solution of Eq. \eqref{eq:theta_simplified_3} can be straightforwardly expressed in the form
\begin{equation}\label{eq:theta_solution}
\vartheta(\bm{q},t) = e^{S(\bm{q},t)}\left[\vartheta(\bm{q},0)+\int_{0}^{t}{\rm d}t'\,e^{-S(\bm{q},t')}f(\bm{q},t')\right]\;,
\end{equation}
where
\begin{equation}
S(\bm{q},t) = - \int_{0}^{t}{\rm d}t'\,\mathcal{L}(\bm{q},t')\;.	
\end{equation}
Without loss of generality, we can choose the initial condition $\vartheta(\bm{r},0)=0$, so that the first term in \eqref{eq:theta_solution} vanishes. The second term can be used to compute the orientational structure factor, yielding:
\begin{equation}
\left\langle |\vartheta(\bm{q},t)|^{2}\right\rangle	
= \frac{2k_{B}T}{\gamma}\,e^{2S(\bm{q},t)}\int_{0}^{t}{\rm d}t'\,e^{-2S(\bm{q},t')}\;.
\end{equation}
In practice, it is more convenient to swap the order of the integrals over $t$ and $\bm{q}$ in Eq. \eqref{eq:q_integral_1} and take advantage of the integration formula for multivariate Gaussian integrals:
\begin{equation}\label{eq:gaussian_integral}
\int_{\mathbb{R}^{d}} {\rm d}^{d}q\,e^{-\frac{1}{2}\bm{q}\cdot\bm{M}\cdot\bm{q}+i\bm{q}\cdot\bm{r}} = \sqrt{\frac{(2\pi)^{d}}{\det\bm{M}}}\,e^{-\frac{1}{2}\bm{r}\cdot\bm{M}^{-1}\cdot\bm{r}}\;.
\end{equation}
with $\bm{M}$ a $d \times d$ matrix of coefficients independent on $\bm{q}$. Notice that, unlike in Eqs. \eqref{eq:q_integral_1}, here the integration is extended over the whole $d-$dimensional real space. This leads to divergences that, nevertheless, cancel out in the connected correlation function upon introducing a suitable short-distance cut-off. To illustrate this strategy, let us consider again the $\dot{\epsilon}=0$ case. The equal-time two-point correlation function can be expressed as:
\begin{equation}
\langle \vartheta(\bm{r},t)\vartheta(\bm{0},t) \rangle 
= \frac{2k_{B}T}{\Ge} \int_{0}^{t} {\rm d}t' \int_{\mathbb{R}^{2}} \frac{d^{2}q}{(2\pi)^{2}}\,e^{-\frac{1}{2}\bm{q}\cdot\bm{M}\cdot\bm{q}+i\bm{q}\cdot\bm{r}}\;
\end{equation}
with $\bm{M}=4\De(t-t')\mathbb{1}$. Calculating the integral over $\bm{q}$ yields:
\begin{align}
\langle \vartheta(\bm{r},t)\vartheta(\bm{0},t) \rangle 
&= \frac{2k_{B}T}{\Ge} \int_{0}^{t}\frac{{\rm d}t'}{2\pi}\,\frac{e^{-\frac{|\bm{r}|^{2}}{8\De(t-t')}}}{4\De(t-t')} \notag \\[10pt]
&= - \frac{k_{B}T}{4 \pi K} \Ei\left(-\frac{|\bm{r}|^{2}}{8\De t}\right)\;,
\end{align}
Now, in the limit of $t\rightarrow\infty$, the exponential integral diverges logarithmically:
\begin{equation}\label{eq:ei_expansion}
\Ei(\pm z) = \gamma_{\rm EM}+\log z \pm z + \mathcal{O}(z^{2})\;,\qquad 0 < z \ll 1\;, \notag
\end{equation}
with $\gamma_{\rm EM}$ the Euler-Mascheroni constant. This singular behavior can be regularized by approximating $\langle |\vartheta(\bm{0},t)|^{2} \rangle\approx\langle \vartheta(\bm{a},t)\vartheta(\bm{0},t)\rangle$, where the vector $\bm{a}= a\bm{e}_{r}$ traces the boundary of a small disk-shaped region around the origin. Then, using the expansion of the exponential integral, we can express the connected correlation function in the standard form in the long time limit. This gives:
\begin{align}\label{eq:g_log}
g(\bm{r})
&\approx -\frac{k_{B}T}{4 \pi K}\left[\Ei\left(-\frac{a^{2}}{8\De t}\right)-\Ei\left(-\frac{|\bm{r}|^{2}}{8\De t}\right)\right] \notag \\[10pt]
&\xrightarrow{t\rightarrow\infty} \frac{k_{B}T}{2\pi K}\,\log\frac{|\bm{r}|}{a}\;,
\end{align}
consistent with the equilibrium result, Eq. \eqref{eq:g_no_flow}. 

Now, for $\dot{\epsilon}\ne 0$, carrying out the integral over $\bm{q}$ yields:
\begin{equation}\label{eq:t_integral}
\langle \vartheta(\bm{r},t)\vartheta(\bm{0},t) \rangle 
= \frac{2k_{B}T}{\Ge} \lim_{t\rightarrow\infty} \int_{0}^{t}\frac{{\rm d}t'}{2\pi}\,\frac{e^{-\frac{1}{2}\bm{r}\cdot\bm{M}^{-1}\cdot\bm{r}}}{\sqrt{\det\bm{M}}}\;,
\end{equation}
where the $\bm{M}$ matrix is given by:
\[
\bm{M} = 4\De\left[
\begin{array}{cc}
t-t' + \frac{1}{3}\dot{\epsilon}^{2}\left(t^3-t'^{3}\right) & -\frac{1}{2}\dot{\epsilon}\left(t^{2}-t'^{2}\right) \\[10pt]
-\frac{1}{2}\dot{\epsilon}\left(t^{2}-t'^{2}\right) & t-t'
\end{array}
\right]\;.
\]
Reintroducing the original coordinates, via Eq. \eqref{eq:galilean}, yields the equal time correlation function in the following integral form:
\begin{multline}
\langle \vartheta(\bm{r},t)\vartheta(\bm{0},t)\rangle =
\frac{k_{B}T}{2\pi K} \int_{0}^{t}
\frac{{\rm d}\Delta t}{ \Delta t \, \sqrt{ 4 + \frac{1}{3}\dot{\epsilon}^{2}\Delta t^{2} } } \\
\exp\left[
-\frac{x^{2}-\dot{\epsilon}\Delta t\,xy + \left(1+\frac{1}{3}\dot{\epsilon}^{2}\Delta t^{2}\right)y^{2}}{2\mathcal{D} \Delta t \left(4+\frac{1}{3}\dot{\epsilon}^{2}\Delta t^{2}\right)}
\right]\;,
\end{multline}
where $\Delta t=t-t'$. Finally, taking $\tau=\dot{\epsilon}\Delta t$, switching to polar coordinates, and taking the limit  $t\rightarrow\infty$, allows one to express the steady state connected correlation function in the form given by Eq. \eqref{eq:g}.

\subsection{Derivation of Eq. \eqref{eq:g_limit}}

As the function $\mathcal{G}(\tau,\phi)$ in Eq. \eqref{eq:phi} approximately scales like $\mathcal{G}(\tau,\phi)\sim 1/\tau$, the exponential factor $\exp[-\mathcal{G}(\tau,\phi)z^{2}]$, with $z\ll 1$, affects the magnitude integrand only for $\tau \ll 1$ and rapidly plateaus to one for $\tau \gg 1$. Taking advantage of this, one can approximate:
\begin{align}\label{eq:integral_split}
\int_{0}^{\infty} {\rm d}\tau\,\frac{e^{-\mathcal{G}(\tau,\phi)z^{2}}}{\tau\sqrt{4+\frac{1}{3}\tau^{2}}} 
&\approx 
\int_{0}^{1}{\rm d}\tau\,
\frac{e^{-\frac{z^{2}}{8\tau}}}{2\tau}+\int_{1}^{\infty}\frac{{\rm d}\tau}{\tau\sqrt{4+\frac{1}{3}\tau^{2}}} \notag \\
&=-\frac{1}{2}\Ei\left(-\frac{z^{2}}{8}\right)+\frac{1}{2}\arcsinh 2\sqrt{3}\;.
\end{align}
This approximation can always be applied to the first integral at the right-hand side Eq. \eqref{eq:g}, since the cut-off radius $a$ is a microscopic length scale, and, for $r \ll \ell_{\rm s}$, to the second integral as well. Using again the expansion of the exponential integral, this readily yields the usual logarithmic dependence, Eq. \eqref{eq:g_log}, in the short distance limit. Similarly, for $r \gg \ell_{\rm s}$ the second integral at the right-hand side of Eq. \eqref{eq:g} vanishes, whereas the first integral yields again Eq. \eqref{eq:integral_split} with $z=a/\ell_{\rm s}$, from which one recovers Eq. \eqref{eq:g_limit}.

\end{document}